\documentclass[journal,10pt,twocolumn,]{IEEEtran}

\usepackage{cite}
\usepackage{amsmath}
\usepackage{graphicx}
\usepackage{enumitem}
\usepackage{bm}
\usepackage{mathtools}
\usepackage{amsfonts}
\usepackage{stackengine}
\usepackage{xcolor}
\usepackage{float}
\stackMath
\usepackage{multirow} 

\usepackage{relsize}
\usepackage{bm}
\usepackage{soul}
\usepackage{subcaption}
\let\latexchi\chi
\makeatletter
\renewcommand\chi{\@ifnextchar_\sub@chi\latexchi}
\newcommand{\sub@chi}[2]{
	\@ifnextchar^{\subsup@chi{#2}}{\latexchi^{}_{#2}}%
}
\newcommand{\subsup@chi}[3]{
	\latexchi_{#1}^{#3}%
}
\makeatother
\def\mathclap#1{\text{\hbox to 0pt{\hss$\mathsurround=0pt#1$\hss}}}
\newcommand{\raisedchi}{\raisebox{\depth}{\(\chi\)}}

\begin{document}

\title{Physics Assisted Deep Learning for Indoor Imaging using Phaseless Wi-Fi Measurements}

\author{Samruddhi~Deshmukh,
         Amartansh~Dubey,
         Dingfei~Ma,~\IEEEmembership{Graduate Student Members,~IEEE},
         Qifeng~Chen,~\IEEEmembership{Member,~IEEE},
         Ross~Murch,~\IEEEmembership{Fellow,~IEEE}
\thanks{This work was supported by the Hong Kong Research Grants Council with the Collaborative Research Fund C6012-20G.}
\thanks{Samruddhi Deshmukh, Amartansh Dubey and Dingfei Ma are with the Department of Electronic and Computer Engineering, Hong Kong University of Science and Technology (HKUST), Hong Kong (e-mail: ssdeshmukh@connect.ust.hk; adubey@connect.ust.hk;  dma@connect.ust.hk).}
\thanks{Qifeng Chen is with the Department of Computer Science and Engineering and the Department of Electronic and Computer Engineering, Hong Kong University of Science and Technology (HKUST), Hong Kong (e-mail: cqf@ust.hk).}
\thanks{Ross Murch is with the Department of Electronic and Computer Engineering and the Institute of Advanced Study, Hong Kong University of Science and Technology (HKUST), Hong Kong (e-mail: eermurch@ust.hk).}
}

\maketitle

\begin{abstract}
A physics assisted deep learning framework to perform accurate indoor imaging using phaseless Wi-Fi measurements is proposed. It is able to image objects that are large (compared to wavelength) and have high permittivity values, that existing radio frequency (RF) inverse scattering techniques find very challenging, making it suitable for indoor RF imaging. The technique utilizes a Rytov based inverse scattering model with a deep learning framework. The inverse scattering model is based on an extended Rytov approximation (xRA) that pre-reconstructs the RF measurements. Under strong scattering conditions, this pre-reconstruction is related to the actual permittivity profile by a non-linear function, which is learned by a modified U-Net model to obtain the permittivity profile of the object. Thus, our proposed approach not only reconstructs the shape of objects, but also estimates their permittivity values accurately. We demonstrate its imaging performance using simulations as well as experimental results in an actual indoor environment using 2.4 GHz Wi-Fi phaseless measurements. For incident wavelength $\lambda_0$, the proposed framework can reconstruct objects with relative permittivity as high as 77 and electrical size as large as $40 \lambda$, where $\lambda =\lambda_0/\sqrt{77}$. This is in contrast to existing phaseless imaging techniques which cannot reconstruct permittivity values beyond 3 or 4. Thus, our proposed method is the first inverse scattering-based deep learning framework which can image large scatterers with high permittivity and achieve accurate indoor RF imaging using phaseless Wi-Fi measurements. 
\end{abstract}


%
\IEEEpeerreviewmaketitle

\section{Introduction}
\label{Sec_intro}
 Inverse scattering problems (ISPs) have been utilized in a wide range of research and commercial applications including non-destructive testing, medical imaging, gas and oil exploration \cite{9201415, murch1990inverse, TCI4, TCI5, chen2020review, TCI2}. Modeling the scattering of waves to infer the shape and constitution of a scatterer or source has provided many breakthroughs in science and technology including Rutherford's iconic scattering experiment to determine the structure of an atom.  

Solving ISPs in their exact form (without any approximations) can provide accurate reconstruction of objects. A common form of ISP is obtained from the Lippmann Schwinger equation which can be derived from Maxwell's wave equations and can model electromagnetic wave-matter interaction \cite{chen2018computational}. However, such ISPs in their original form can be highly non-linear and severely ill-posed due to several physical constraints. For example, the extent of the non-linearity depends on factors such as the material constitution of the object, wavelength of the probing/incident wave $\lambda_0$ and the size of the Domain of Interest (DoI).  The extent of ill-posedness depends on factors such as the number of measurements (number of knowns), size of the DoI to be imaged (number of unknowns) and the type of noise/distortion in the measurements. 

Due to the difficulty of obtaining solutions to ISPs they often have a limited range of applicability. Challenging scenarios for ISPs include objects with high permittivity value and DoI and scatterers with sizes comparable or greater than the wavelength $\lambda_0$ of the probing waves. To the best of our knowledge, almost all linear as well as non-linear state-of-the-art techniques fail when the scatterer is both large in size and has a high permittivity value (or do not show results for such cases) \cite{TCI1, chen2018computational, Xudongchen, chen2020review}. Most linear techniques are predicated on scenarios with weak scattering where permittivity deviates only slightly from free-space. Non-linear techniques can moderately extend this range (to relative permittivity $\epsilon_r<3$ or $4$) at the expense of higher computational cost and higher sensitivity to measurement errors and rely on controlled anechoic chamber experiments \cite{TCI1, chen2018computational, Xudongchen, chen2020review}. Also, these techniques only demonstrate results for scatterers smaller than the incident wavelength $\lambda_0$.

Due to the limited range of applicability of most existing inverse scattering techniques they are not applicable to scenarios such as indoor imaging using Wi-Fi signals. In these applications the size of the DoI and scatterer can be as high as 10$\times$ or even 100$\times$ the incident wavelength $\lambda_0$ and the permittivity of the scatterers can be as high as $\epsilon_r>50$ (for objects containing water such as the human body). This vast range in size and permittivity values of the scatterers is well beyond the validity range of any existing inverse scattering technique. Furthermore, applications such as indoor imaging need a robust and practical measurement system that does not need high precision measurements of phase and synchronization among different nodes, which makes techniques that utilize phaseless measurements more desirable \cite{Xudongchen}.

Empirical techniques such as radio tomographic imaging (RTI) \cite{Patwari2010, WangRTI2016,TomicRTI2015, 1Patwari2013, depatla2015x, Patwari2017, Patwari2015, Mostofi2014} try to partially address the challenge of indoor RF imaging. These reconstruct a binary image of the scatterer indicating its \textit{presence} or \textit{absence}, but provide no reconstruction of its permittivity value and hence fail in tasks such as imaging multiple objects with different permittivity values. These techniques also do not have a solid theoretical formulation and only provide experimental results making them empirical in nature. 

In this work our goal is to provide a phaseless inverse scattering framework for achieving high resolution indoor RF imaging using Wi-Fi RSSI measurements. To achieve this, we use a physics assisted deep learning approach. The physics assisted learning approach \cite{jin2017deep} involves obtaining an initial guess of the permittivity profile from a physics based inverse model and using this as input to a neural network (NN) to learn the final permittivity profile. This is in contrast to having a NN learn a direct mapping from the measurements to the permittivity profile (in a ``black box"/direct learning approach). Thus, we are incorporating the information of the physics based model in the NN through the initial guess. In this approach, the selection of the inverse model is as important as the choice of the NN architecture.

\subsection{Contributions}
\label{sec_contri}
Specific key contributions of this work include:
\begin{enumerate}
	\item \textbf{Overall Framework: }We develop a physics assisted deep learning framework which can perform high resolution indoor imaging using phaseless (RSSI) Wi-Fi measurements. 
	\item \textbf{Physics based model}: We propose a phaseless inverse scattering physics based assistance model using an extended Rytov approximation, which we denote as xRA. This is used as the inverse model to generate an initial reconstruction. The novelty of this approach is that the initial reconstruction provides both a real and imaginary image output, both of which encode information about permittivity in the form of a non-linear function. A deep learning framework then takes both images to learn the final permittivity profile from this non-linear relationship. 
	\item \textbf{Deep Learning Framework}: We use a modified U-Net architecture, which is an encoder-decoder based Convolutional Neural Network (CNN) with skip connections between the encoder and decoder layers. The modification includes adding additional filters to the skip connections to reduce the semantic gap between outputs of the encoder and decoder layers. The input to the network consists of two channels, each representing the real and imaginary parts of the initial reconstruction generated by the xRA model and the output of this network is the final permittivity profile of the imaging region.
\end{enumerate}

In xRA a Wentzel–Kramers–Brillouin (WKB) approximation is utilized to approximate the field inside the object to show that both the real and imaginary reconstructions using the Rytov approximation (RA) contain useful information about the object's permittivity. Further, the use of a phaseless formulation for xRA allows it to be utilized with a low cost measurement system. In addition, the background subtraction framework is incorporated in xRA to remove multipath scattering from background clutter. 

To the best of our knowledge, these modifications can only be made to RA, thus making it a unique choice in terms of assisting the NN for use in indoor imaging. For example, most recent phaseless inverse scattering methods \cite{Xudongchen} cannot be reformulated for background subtraction since the change in magnitude of the electric field is not linearly related to the change in the contrast function. Incorporating background subtraction in the model is a key for indoor imaging since it cancels out the scattered fields arising from the stationary background including the ceiling, floor and other clutter.

The aforementioned contributions lead to a practically feasible physics assisted deep learning framework which is not only general, but also provides the following key advantages over existing methods: 
\begin{enumerate}
	\item \textbf{Range of validity} (in terms of scatterer permittivity and size): Our proposed technique outperforms the validity range achieved by other phaseless inverse scattering approaches by a significant margin. We show results for extremely high permittivity values and cover the
	typical permittivity range needed for indoor imaging ($1< \epsilon_r < 77$). Even the best existing phaseless inverse scattering techniques only performs for $\epsilon_r < 3$. Our technique also perform for scatterers much larger than the incident wavelength $\lambda_0$ (electrical size $> 40 \lambda$, where $\lambda = \lambda_0/\sqrt{\epsilon_r}$) whereas most recent phaseless methods start to break down as scatterer size approaches $\lambda_0$.   	
	\item \textbf{Generalization from Simulation to Experiments:} The proposed framework, which is trained on 2D simulation data, provides highly accurate results for experimental data obtained from a 3D indoor environment containing measurement error and multipath reflections (due to ceiling, floor and other clutter in the DoI). This removes the need to collect experimental data for training the NN, which makes our framework practical for indoor RF imaging. This generalization can be partly attributed to the robustness of our xRA model in physical experiments. Most of the existing work \cite{ Xudongchen} is only generalized in highly controlled experiments (e.g. in anechoic chambers) where the  experimental data is accurate with low noise.	
	\item \textbf{Computational Complexity}: The xRA model is linear in its formulation. This makes the generation of a simulation-based training dataset fast and also reduces the inference time of the technique.
\end{enumerate}

\subsection{Organization and Notation}
\label{Sec_Org}

The paper is organized as follows. Section \ref{Sec_PS} describes the problem setup followed by Section \ref{Sec_RAf} explaining the physics based inverse model used to assist the deep learning framework. Section \ref{Sec_DLframework} describes the deep learning framework used and its suitability for the inverse scattering problem. This is followed by Section \ref{Sec_NumericalResults} showing results using simulation data and Section \ref{Sec_ExperimentalResults} showing results using experimental data obtained from an indoor region.

Throughout the remainder of the paper, we use lowercase boldfaced letters (\textbf{x}) to represent position vectors, single overline $(\overline{X})$ to represent vectors and double overline $(\overline{\overline{X}})$ to represent matrices.

\section{Problem Setup}
\label{Sec_PS}

\begin{figure}[h]
	\centering
	\includegraphics[width=3in]{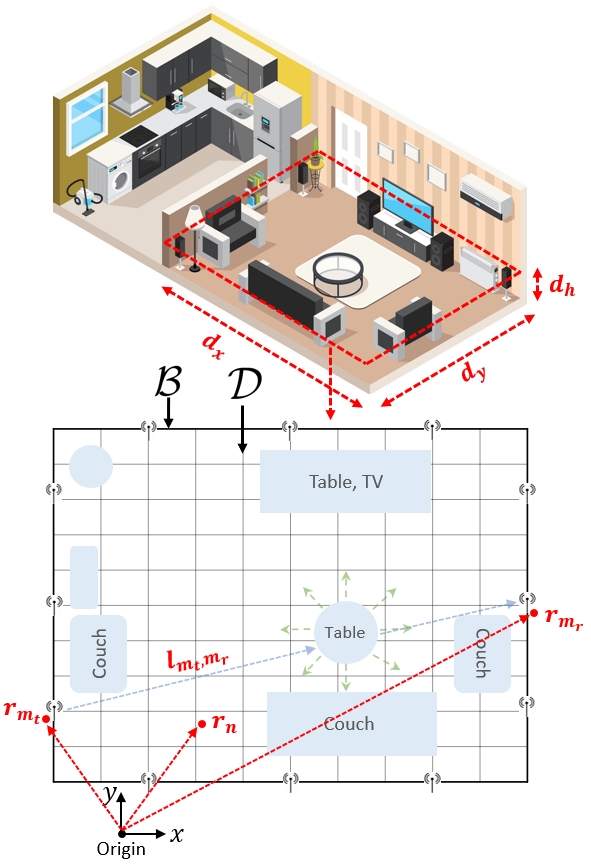}
	\caption{The DoI with wireless transceiver nodes at its boundary $ {\mathcal{B}}$. The transmitter $m_t$ and receiver $m_r$ are located at ${\bf r}_{m_t} \in {\mathcal{B}}$ and ${\bf r}_{m_r} \in {\mathcal{B}}$ respectively. The location of the $n^{th}$ grid in the DoI is denoted as ${\bf r}_n  \in {\mathcal{D}}$.}
	\label{RTInetwork2}
\end{figure}

Let ${\mathcal{D}}\subset \mathbb{R}^{2}$ be the DoI  of dimensions $d_x \times d_y$ m$^2$ that is situated in an indoor region of a building as shown in Fig. \ref{RTInetwork2}.  Measurements are obtained using standard Wi-Fi transceivers operating at 2.4 GHz. These wireless transceiver nodes are placed at the boundary of the DoI (denoted as $\mathcal{B} \subset \mathbb{R}^{2}$). They can transmit and receive signals to acquire RSSI measurements of the links between nodes. There are $M$ nodes in total, making the total number of wireless links equal to $L=M(M-1)/2$. Since nodes cannot usually transmit and receive at the same time, the $L$ links do not include reciprocal links or self measurements. The location of the $m^{th}$ measurement node is denoted as ${\bf r}_m \in \mathcal{B}$ and the link between transmitting node $m_t$ (at ${\bf r}_{m_t}$) and receiving node $m_r$ (at ${\bf r}_{m_r}$) is denoted by $l_{m_t,m_r}$. Throughout the remainder of this work, the subscripts $m_t$ and $m_r$ are used to represent the transmitter and receiver respectively for all relevant quantities. 

We consider the DoI to be a planar cross-section parallel to the floor at a height $d_h$. We can therefore approximate the environment as a two-dimensional (2D) electromagnetic problem. This 2D approximation is valid for the assumption that there is negligible scattering from the floor, ceiling and other clutter so that the scattering is mainly from objects kept in the DoI cross-section. We achieve this by using directive antennas at the nodes and also using the background subtraction framework of xRA as explained later in Section \ref{Sec_TBS}. Further details of the measurement system are provided later in Section \ref{Sec_ExperimentalSetup}.

For imaging, we discretize the DoI into $N = n_x \times n_y$ small rectangular grids (of dimension $\Delta d_x \times \Delta d_y$). The convention in inverse scattering is to make these grids much smaller than the incident wavelength ($\Delta d_x, \Delta d_y \ll \lambda_0$) \cite{chen2018computational}. The location of the $n^{th}$ grid is denoted by ${\bf r}_n$ (see Fig. \ref{RTInetwork2}). Our goal is to estimate the permittivity for all $N$ grids and generate a 2D permittivity profile of the DoI. In essence, we aim to solve for $N$ unknowns using $L=M(M-1)/2$ measurements.

\section{Physics based Assistance Model}
\label{Sec_RAf}
In this section we detail the xRA model that assists the deep learning framework. The basic RA model with phaseless data is first introduced, followed by its extended form. In addition background subtraction is introduced which is a key technique for use to experimentally remove background scattering or clutter. Finally, the discretization and regularization of the problem is considered. 
\subsection{Rytov Phaseless Approximation}
Let the incident field from the $m_t^{th}$ node, at any point $\bm{r}_n$ inside the DoI in the absence of scatterers, be denoted as $E_{m_t}^i(\bm{r}_n)$. In the presence of scatterers, the total field at any point $\bm{r}_n$ is denoted as $E_{m_t}(\bm{r}_n)$. Conventionally, the total field is expressed as the sum of the incident and scattered fields and written as $E_{m_t} = E^i_{m_t} + E^s_{m_t}$. For RA, the expression for the total field is considered in a way that can incorporate the wavefront curvature as the field propagates through a medium. Therefore, the total field is expressed as
\begin{equation}
	\label{Eq_tot_field}
	E_{m_t}(\bm{r_n}) = E^i_{m_t}(\bm{r_n})e^{jk_0 \phi^s_{m_t}(\bm{r}_n)},
\end{equation}
where $k_0$ is the wave number and $\phi^s_{m_t}$ is a complex function that can encapsulate the amplitude and phase change in the incident field as it propagates through the media. 

The incident field ${E^i_{m_t}(\bm{r_n})}$ satisfies the homogeneous Helmholtz wave equation for free space (\ref{Eq_wave_equation}a) and the total field ${E_{m_t}(\bm{r_n})}$ satisfies the inhomogeneous Helmholtz wave equation (\ref{Eq_wave_equation}b), 
\begin{subequations}
	\label{Eq_wave_equation}
	\begin{align}
	(\nabla^2 + k_0^2) E^i_{m_t}(\bm{r}_n)& = 0, \\
	(\nabla^2 + k_0^2 \nu^2(\bm{r}_n)) E_{m_t}(\bm{r}_n) & = 0,
	\end{align}
\end{subequations}
where $\nu$ is the refractive index, which is related to the relative permittivity $\epsilon_r$ as $\epsilon_r = \nu^2 $. Unless mentioned otherwise, we only consider lossless materials so that the relative permittivity can be written as $\epsilon_r = \epsilon_R + j\epsilon_I$ where $\epsilon_I = 0$. 

By substituting (\ref{Eq_tot_field}) into (\ref{Eq_wave_equation}b) and subtracting it from (\ref{Eq_wave_equation}a) we obtain the non-linear differential equation which can be written in an integral form (which we call the Rytov integral, RI) to relate the received field with the scatterer profile as
\begin{subequations}
	\label{Eq_RAintegral}
	\begin{align}
		 \tilde{E}_{m_t}(\bm{r}_{m_r}) = & k^2  \int_{{\mathcal{D}}}   g(\bm{r}_{m_r}, \bm{r}_n) \biggl[\nu(\bm{r}_n)^2-1 \nonumber \\
		 &- \nabla \phi^s_{m_t}(\bm{r}_n) \cdot \nabla \phi^s_{m_t}(\bm{r}_n)\biggr] E^i_{m_t}(\bm{r}_n) d\bm{r}^2,
	\end{align}
	\begin{align}
		 \tilde{E}_{m_t}(\bm{r}_{m_r}) &= E^i_{m_t}(\bm{r}_{m_r}) \ln\frac{E_{m_t}(\bm{r}_{m_r})}{E^i_{m_t}(\bm{r}_{m_r})},
	\end{align}
\end{subequations}
where $\bm{r}_{m_t}$ and $\bm{r}_{m_r}$ are the locations of the transmitting and receiving nodes at the measurement boundary, $\bm{r}_{m_t}, \bm{r}_{m_r} \in \mathcal{B}$, and $\bm{r}_n$ is a point in the DoI, $\bm{r}_n \in \mathcal{D}$. We can write (\ref{Eq_RAintegral}a) in exponential form as 
\begin{equation}
	\label{Eq_total_field_exp}
	\begin{aligned}
		\frac{E_{m_t}(\bm{r}_{m_r})}{E^i_{m_t}(\bm{r}_{m_r})} = & \exp\biggl(  \frac{k^2}{E^i_{m_t}(\bm{r}_{m_r})} \int_{{\mathcal{D}}}  g(\bm{r}_{m_r}, \bm{r}_n) \biggl[\nu(\bm{r}_n)^2 - 1  \\
		&- \nabla \phi^s_{m_t}(\bm{r}_n) \cdot \nabla \phi^s_{m_t}(\bm{r}_n)\biggr] E^i_{m_t}(\bm{r}_n) d\bm{r}^2\biggr).
	\end{aligned}
\end{equation}

Multiplying (\ref{Eq_total_field_exp}) by its conjugate and taking $\log_{10}$ on both sides gives us the phaseless form of (\ref{Eq_total_field_exp}) in terms of the total and incident powers (in dB) as 
\begin{subequations}
	\label{Eq_total_power}
	\begin{align}
		& P_{m_t}  (\bm{r}_{m_r}) [\text{\small dB}]  =  P_i  (\bm{r}_{m_r}) [\text{\small dB}] + \nonumber \\ & C_0  \cdot \operatorname{Re}\biggl(\frac{k^2}{E^i_{m_t}(\bm{r}_{m_r})} \int_{{\mathcal{D}}}  g(\bm{r}_{m_r}, \bm{r}_n) \chi_{\text{RI}}(\bm{r}_n) E_i(\bm{r}_n) d\bm{r}^2\biggr),
	\end{align}
	\begin{align}
		\chi_{\text{RI}}(\bm{r}_n) & = \nu(\bm{r}_n)^2-1 - \nabla \phi^s_{m_t}(\bm{r}_n) \cdot \nabla \phi^s_{m_t}(\bm{r}_n),
	\end{align}
\end{subequations}
where $C_0 = 20\log_{10}e$ is a constant. Solving (\ref{Eq_total_power}) as an inverse problem requires estimating two variables $\nu(\bm{r}_n)$ and $\nabla \phi^s(\bm{r}_n)$ from one known $P_{m_t}  (\bm{r}_{m_r})$ making it a difficult non-linear and ill-posed problem. 

RA is derived by ignoring the term $\nabla \phi^s \cdot \nabla \phi^s $ which results in the contrast function
\begin{equation}
	\label{Eq_RAcontrast}
	\begin{aligned}
		\chi_{\text{RA}} (\bm{r}_n) = \nu(\bm{r}_n)^2-1 = \epsilon_r(\bm{r}_n) -1.
	\end{aligned}
\end{equation}
However, the assumption that $\nabla \phi^s \cdot \nabla \phi^s $ can be ignored is valid only under weak scattering conditions, i.e. when $\nabla \phi^s \cdot \nabla \phi^s $ is small, which is why it is not always useful for practical applications. Therefore to use (\ref{Eq_total_power}) for practical imaging scenarios under strong scattering conditions, we need to estimate $\nabla \phi^s \cdot \nabla \phi^s $. In the next subsection, we estimate this term using the WKB approximation to obtain an approximate expression for (\ref{Eq_total_power}) which can be used under strong scattering conditions.

\subsection{Extended Rytov Approximation}
\label{Sec_WKB}
For approximating $\nabla \phi^s \cdot \nabla \phi^s $, the WKB approximation \cite{Chew1999} has been utilized to extend the validity of the Rytov approximation \cite{bates1976extended, murch1990inverse}. The reason why we utilize xRA as the model for assisting deep learning is that the model shows that both the real and imaginary components of the reconstruction contain accurate information about $\epsilon_r$.

The WKB approximation is considered an appropriate approximation for high frequency scenarios including indoor imaging using Wi-Fi, where most of the objects (or inhomogenities) are larger than the incident wavelength $\lambda_0$ (for 2.4 GHz and 5 GHz Wi-Fi, $\lambda_0 = 12.5$ cm and $6.25$ cm respectively). 

Under the high frequency assumption, the total field inside a scatterer can be approximated using WKB approximation as
\begin{equation}
	\label{Eq_WKB}
	\begin{aligned}
		E_{m_t}(\bm{r}_n) =  \frac{E_0}{\sqrt{\nu(\bm{r}_n)}} \exp({j k_0 \int\displaylimits_{\mathclap{\text{along $\bm{\hat{k}}$}}} \nu(\bm{r}_n) \bm{\hat{k}} \cdot \bm{dr}}),
	\end{aligned}
\end{equation}
where $\bm{\hat{k}}$ is the direction normal to the wavefront of the total field inside the scatterer and $\bm{dr} = \bm{\hat{k}} dr$ is the differential path. This high frequency approximation to the total field can be equated to total field in RI (\ref{Eq_tot_field}) as
\begin{equation}
	\label{Eq_nabla_phi}
	\begin{aligned}
		E^i_{m_t}(\bm{r}_n)  e^{jk_0 \phi^s_{m_t}(\bm{r}_n)} = \frac{E_0}{\sqrt{\nu(\bm{r}_n)}} \exp({j k_0 \int\displaylimits_{\mathclap{\text{along $\bm{\hat{k}}$}}} \nu(\bm{r}_n) \bm{\hat{k}} \cdot \bm{dr}}),
	\end{aligned}
\end{equation}
where the incident field can be written as $E_i(\bm{r}_n) = E_0  e^{j k_0 \bm{\hat{k}_i \cdot \bm{r}_n}}$. From (\ref{Eq_nabla_phi}) we can estimate the required term as
\begin{equation}
	\label{Eq_nabla_phi3}
	\begin{aligned}
		\nabla  \phi^s_{m_t} & (\bm{r}_n) \cdot \nabla \phi^s_{m_t}(\bm{r}_n) =  \bigl[\nu^2 + 1 - 2\nu \bm{\hat{k}} \cdot \bm{\hat{k}_i}\bigr] \\ 
		&- \frac{1}{4 k_0^2} \nabla \ln (\nu) \cdot \nabla \ln (\nu) + \frac{j}{k_0} \nabla \ln({\nu}) \bigl[\nu \bm{\hat{k}} - \bm{\hat{k}_i} \bigr],
	\end{aligned}
\end{equation}
where $\nu$ is function of location $\bm{r}_n$ and $\bm{\hat{k}} \cdot \bm{\hat{k}_i} = \cos\theta_s$ is the scattering angle. Substituting $\nabla \phi^s_{m_t} \cdot \nabla \phi^s_{m_t}$ from (\ref{Eq_nabla_phi3}) into (\ref{Eq_RAintegral}b) and expressing $\nu$ in terms of $\epsilon_r$ gives
\begin{equation}
	\label{Eq_newcontrast}
	\begin{aligned}
		\chi_{\text{RI}}(\bm{r}_n) &= \nu(\bm{r}_n)^2-1 -  \nabla \phi_s(\bm{r}_n) \cdot \nabla \phi_s(\bm{r}_n)\\
		& = 2(\sqrt{\epsilon_r} \cos\theta_{s} -1) + \frac{1}{4 k_0^2} \nabla \ln (\sqrt{\epsilon_r}) \cdot \nabla \ln (\sqrt{\epsilon_r}) \\ &   \qquad \qquad \qquad  + j \biggl(\frac{1}{k_0} \nabla \ln({\sqrt{\epsilon_r}}) \bigl[\sqrt{\epsilon_r} \bm{\hat{k}} - \bm{\hat{k}_i}\bigr] \biggr),
	\end{aligned}
\end{equation}
where $\epsilon_r$ is a function of location $\bm{r}_n$. The contrast $\chi_{\text{RI}}$ (\ref{Eq_newcontrast}) can provide more accurate shape reconstruction than $\chi_{\text{RA}}$ (\ref{Eq_RAcontrast}) under strong scattering conditions as $\chi_{\text{RI}}$ does not ignore the $\nabla \phi^s \cdot \nabla \phi^s $ term \cite{dubey2021enhanced}. However, unlike conventional RA in which contrast $\chi_{\text{RA}}$ is a straightforward linear function of the permittivity value $\epsilon_r$, the contrast $\chi_{\text{RI}}$ in (\ref{Eq_newcontrast}) is a non-linear function of $\epsilon_r$ and involves other unknown entities such as the scattering angle ($\theta_s$) which itself is a function of the scatterer's shape and permittivity.

Another key point to note is that for a lossless scatterer ($\epsilon_I=0$), $\chi_{\text{RI}}$ in (\ref{Eq_newcontrast}) still has a non-zero imaginary component (unlike in conventional RA where $\chi_{\text{RA}}$ (\ref{Eq_RAcontrast}) is purely real for the lossless case). Therefore, the real and imaginary components of the contrast $\chi_{\text{RI}}$ can be expressed as non-linear functions of $\epsilon_R$ as 
\begin{equation}
	\label{Eq_newconstrastfinal}
	\begin{aligned}
		&\operatorname{Re}(\chi_{\text{RI}})  = f(\epsilon_R), \\ & \operatorname{Im}(\chi_{\text{RI}}) = g(\epsilon_R),	
	\end{aligned}
\end{equation}
where $f$ and $g$ are non-linear functions which characterize (\ref{Eq_newcontrast}). This shows that there is information about $\epsilon_R$ in both real and imaginary components of $\chi_{\text{RI}}$. This leakage of $\epsilon_R$ into the imaginary component of contrast $\chi_{\text{RI}}$ is termed `cross-talk' and has been observed in previous numerical studies \cite{intes2005multi, sebbah2001waves}.

While the information about $\epsilon_R$ in both $\operatorname{Re}(\chi_{\text{RI}})$ and $\operatorname{Im}(\chi_{\text{RI}})$ may be non-linearly related, it has potential for use with deep learning techniques which can learn these non-linear functions to obtain estimates of $\epsilon_R$. 

The profiles $\operatorname{Re}(\chi_{\text{RI}})$ and $\operatorname{Im}(\chi_{\text{RI}})$ can be easily estimated using the measured power (by solving the linear inverse problem in (\ref{Eq_total_power}). 

Our goal in this paper is to estimate the  $\epsilon_R$ profile of the scatterer using the reconstructed contrast profiles $\operatorname{Re}(\chi_{\text{RI}})$ and $\operatorname{Im}(\chi_{\text{RI}})$. This is difficult since the mapping between them is non-linear, which is why we use deep learning to solve this problem. We use both $\operatorname{Re}(\chi_{\text{RI}})$ and  $\operatorname{Im}(\chi_{\text{RI}})$ as an initial guess and supply these as inputs to the neural network to learn the mapping to $\epsilon_R$.

\subsection{Background Subtraction}
\label{Sec_TBS}
One aspect of indoor imaging is that the obtained measurements also contain clutter due to scattering from objects placed outside the DoI due to multipath reflections originating in the DoI. A useful approach for removing this clutter is background subtraction. The technique is based on utilizing measurements from different time instants so that changes in the indoor environment are imaged. In this section, we extend (\ref{Eq_total_power}) (with the derived contrast $\chi_{\text{RI}}$ in (\ref{Eq_newcontrast})) to incorporate background subtraction. 

Consider an initial time instance $t_0$ at which the contrast profile of the DoI is given by $\chi_{\text{RI}}^{t_0}$ and the total power measured is ${P}_{m_t}^{t_0}(\bm{r}_{m_r})$. Let there be some perturbation in this profile in the duration $\Delta t$ and the contrast profile of the DoI at time $t_0+\Delta t$ be $\chi_{\text{RI}}^{t_0+\Delta t}$ with total power measured being $P_{m_t}^{t_0+\Delta t}(\bm{r}_{m_r})$. Writing (\ref{Eq_total_power}) for both time instances and subtracting them gives
\begin{subequations}
	\label{Eq_PRAV1}
	\begin{align}
		 & \Delta  P_{m_t}  (\bm{r}_{m_r}) [\text{\small dB}] \nonumber \\
		 & = C_0  \cdot \operatorname{Re}\biggl(\frac{k^2}{E^i_{m_t}(\bm{r}_{m_r})} \int_{{\mathcal{D}}}  g(\bm{r}_{m_r}, \bm{r}_n)  \Delta \chi_{\text{RI}}(\bm{r}_n) E_i(\bm{r}_n) d\bm{r}^2\biggr),
	\end{align}
	\begin{align}
		\Delta & P_{m_t}  (\bm{r}_{m_r}) [\text{\small dB}] = (P^{t_0+\Delta t}_{m_t} (\bm{r}_{m_r}) - P^{t_0}_{m_t} (\bm{r}_{m_r}))[\text{\small dB}],
	\end{align}
	\begin{align}
		 &\Delta \chi_{\text{RI}}(\bm{r}_n) = \chi_{\text{RI}}^{t_0+\Delta t}(\bm{r}_n) - \chi_{\text{RI}}^{t_0}(\bm{r}_n),
	\end{align}
\end{subequations}
where $C_0 = 20 \log_{10} e$. Equation (\ref{Eq_PRAV1}) provides the phaseless form of xRA with the background subtraction framework. If the DoI background is free space at $t_0$, there is no need for background subtraction and the background field reduces to the free space incident field $P_{m_t}^i$. Equation (\ref{Eq_PRAV1}) represents the final form of the proposed Extended Rytov approximation (xRA).

\subsection{Discretization and Regularization}
The xRA inverse model proposed in (\ref{Eq_PRAV1}) is for a single wireless link $\textit{l}$ between transmitter $m_t$ and receiver $m_r$. The indoor imaging setup shown in Fig. \ref{RTInetwork2} contains $M$ transceiver nodes, so there are a total of $L = M(M-1)/2$ unique wireless links (ignoring the reciprocal links). We need to use these limited measurements to provide the reconstruction where the number of unknowns ($N$) are greater than the number of measurements ($L$). In this section we describe the process to create the reconstructions. 

We can stack the xRA inverse model (\ref{Eq_PRAV1}) for all $L$ links as 
\begin{equation}
	\label{Eq_discrete1}
	\overline{P} = \operatorname{Re}\big(\overline{\overline{G}} \  {\overline{\raisedchi}_{\text{RI}}} \big),
\end{equation}
where each element of the measurement vector $\overline{P} \in \mathbb{R}^{L\times 1}$ is $p_l= \Delta {P}_{m_t}({\bf r}_{m_r})$ [in dB]. Here, $l=m_t(m_r-\text{ceil}(m/M))$ where $m=1,2,...,M-1$ and for each value of $m_t$, $m_r=1+\text{ceil}(m/M),...,M$ (ceil denotes the ceiling or round up function). The contrast vector ${\overline{\raisedchi}_{\text{RI}}}\in \mathbb{R}^{N\times 1}$ contains elements ${{\raisedchi}_{\text{RI}}}(\bm{r}_n)$ for $n = 1, 2,...,N$. The xRA model matrix ${\overline{{\overline{G}}}} \in \mathbb{R}^{L\times N}$ contains entries ${\overline{{\overline{G}}}}_{l,n} $ given as 
\begin{equation}
	\label{Eq_discrete2}
	{\overline{{\overline{G}}}}_{l,n} = C_0 \frac{k_0^2}{E_{m_t}^i(\bm{r}_{m_r})}\sum_{\forall n}  g(\bm{r}_{m_r}, \bm{r}_n)  \chi_{\text{RI}}(\bm{r}_n) E_{m_t}^i(\bm{r}_n) \Delta a,
\end{equation}
where $\Delta a$ is the area of a single grid and $l$ is related to $m_t$ and $m_r$ as stated above so that
\begin{equation}
	\label{Eq_discrete3}
	\overline{P} = \big[\ \operatorname{Re}({\overline{\overline{G}}}) \ \ \ -\operatorname{Im}({\overline{\overline{G}}}) \ \big]\ \begin{bmatrix}
		\operatorname{Re}({\overline{\raisedchi}_{\text{RI}}}) \\
		\operatorname{Im}({\overline{\raisedchi}_{\text{RI}}})
	\end{bmatrix}.
\end{equation}

Equation (\ref{Eq_discrete3}) is the discrete form of xRA without background subtraction. xRA with background subtraction (\ref{Eq_PRAV1}) can be similarly converted to a discrete form. Note that  (\ref{Eq_discrete3}) requires finding the solution to $2N$ unknowns using $L=M(M-1)/2$ measurements. In indoor imaging, we have a limited number of Wi-Fi nodes. Thus, if the DoI is large (which is usually the case), $N$ will also be large, making (\ref{Eq_discrete3}) severely under-determined ($2 N \gg L$). Therefore, we cannot solve it using conventional least squares formulation, since it does not have a unique solution. We therefore use regularization
to impose constraints on the solution domain to remove the ill-posedness of the problem.

Different regularization techniques such as LASSO \cite{lasso}, Tikhonov regularization \cite{tikhonov} and Total Variation (TV) regularization \cite{tvregularization} can be used. In this paper we use \textit{Tikhonov regularization}, which adds a penalty term to the least squares formulation, resulting in the optimization function
\begin{equation}
	\label{Eq_tikhonovopt}
   \underset{(\overline{\raisedchi} \in \mathbb{R}^{2 N\times 1})}{\text{min }} \ \ \frac{1}{2} ||\overline{\overline{\mathcal{G}}}\  \overline{\raisedchi} - \overline{P}||^2 + \alpha ||\overline{\overline{Q}} \overline{\raisedchi}||^2,
\end{equation}
where $\overline{\overline{Q}}$ is the Tikhonov matrix and $\alpha$ is the regularization parameter. $\overline{\overline{\mathcal{G}}}= 
 \big[\operatorname{Re}({\overline{\overline{G}}}) \ -\operatorname{Im}({\overline{\overline{G}}})\big]$
contains the real and imaginary components of the xRA model matrix $G$ concatenated horizontally and
$\overline{\raisedchi} = \big[(\operatorname{Re}({\overline{\raisedchi}_{\text{RI}}}))^T \  (	\operatorname{Im}({\overline{\raisedchi}_{\text{RI}}}))^T \big]^T$ is the variable vector containing Re($\overline{\chi}_{\text{RI}}$) and Im($\overline{\chi}_{\text{RI}}$) concatenated vertically. The Tikhonov matrix $\overline{\overline{Q}}$ enforces certain desired properties on the solution. In this work, we use a difference matrix approximating the derivative operator as the Tikhonov matrix. Since the image is 2D, we include the difference operators in both horizontal and vertical directions. This form of Tikhonov regularization is known as \textit{H1} regularization.

If $D_X$ and $D_Y$ are the difference operators in the horizontal and vertical directions respectively, the solution to (\ref{Eq_tikhonovopt}) is given as 
\begin{equation}
	\label{Eq_h1sol}
\begin{aligned}
	\overline{\raisedchi} = \begin{bmatrix}
	\operatorname{Re}({\overline{\raisedchi}_{\text{RI}}}) \\
	\operatorname{Im}({\overline{\raisedchi}_{\text{RI}}})\end{bmatrix} =& (\overline{\overline{\mathcal{G}}}^T \overline{\overline{\mathcal{G}}} + \alpha (D^T_X D_X + D^T_Y D_Y))^{-1} \overline{\overline{\mathcal{G}}}^T\  \overline{P} \\ & = \overline{\overline{\Pi}} \ \overline{P}.
\end{aligned}
\end{equation}
There are more advanced regularization techniques such as LASSO and Total Variation that enforce better sparsity and smoothness priors as compared to Tikhonov regularization which can lead to better reconstructions. However, we have chosen to use Tikhonov regularization due to three main reasons as follows:
\begin{enumerate}
	\item It is smooth and strongly convex in its formulation, so it has an analytical solution. This has lower computational complexity as compared to solving an optimization problem with a non-smooth objective. 
	\item We can think of the solution of Tikhonov regularization in (\ref{Eq_h1sol}) as the transformation of the measurement vector $\overline{P}$ by projection on $\overline{\overline{\Pi}}$. The transformation matrix $\overline{\overline{\Pi}}$ does not depend on measurements, so it can be pre-calculated and applied to measurements in real time. This makes the process of generating 100s or 1000s of reconstructions for training much faster, and also reduces inference time.
	\item Techniques like LASSO and TV regularization enforce sparsity and continuity in reconstructions by shrinking coefficients. This can lead to the loss of information that xRA preserves if high regularization is applied, which in turn can affect the neural network's ability to learn the mapping between Re($\chi_{\text{RI}}$), Im($\chi_{\text{RI}}$) and $\epsilon_R$ to predict $\epsilon_R$ values in the output accurately.
\end{enumerate}

Solving (\ref{Eq_h1sol}) gives the real and imaginary part of the contrast profile $\operatorname{Re}(\chi_{\text{RI}})$ and $\operatorname{Im}(\chi_{\text{RI}})$ respectively, which are related to the desired permittivity profile $\epsilon_R$ via a non-linear differential equation as shown in (\ref{Eq_newcontrast}). In the next section, we provide details of the deep learning framework we use to model this non-linear function. 

\section{Deep Learning Framework}
\label{Sec_DLframework}

\begin{figure*}
	\centering
	\includegraphics[scale=0.55]{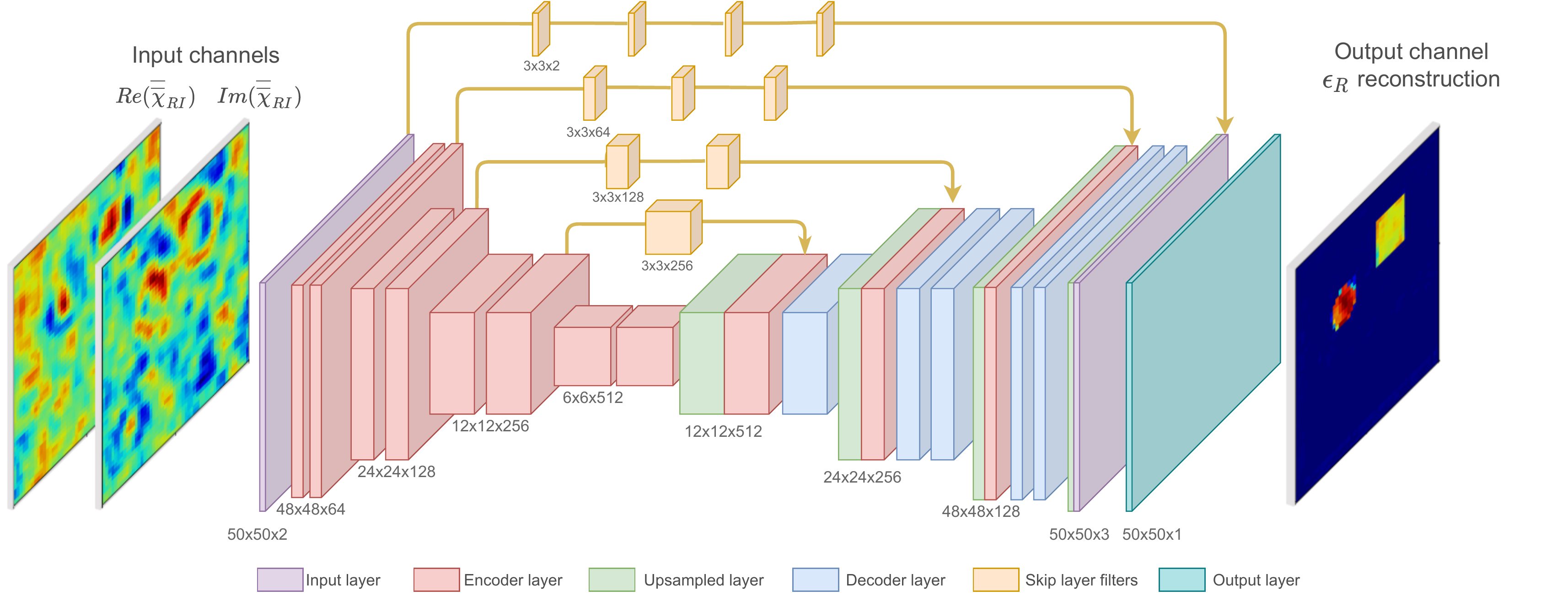}
	\caption{Modified U-Net architecture. All convolutional layers are followed by batch normalization and ReLU activation.}
	\label{unet}
\end{figure*}
Reconstructing the permittivity profile $\epsilon_R$ of the DoI given the real and imaginary parts of the contrast ($\operatorname{Re}(\chi_{\text{RI}})$ and $\operatorname{Im}(\chi_{\text{RI}})$) can be treated as a dense prediction problem. Its objective is to estimate the accurate location and shape of the scatterers present in the DoI, and to obtain accurate permittivity values for each location in the DoI. As shown in (\ref{Eq_newcontrast}), $\epsilon_R$ is related to $\operatorname{Re}(\chi_{\text{RI}})$ and $\operatorname{Im}(\chi_{\text{RI}})$ through a non-linear equation. To model this relationship, we use a modified version of the classic U-Net architecture \cite{unetoriginal}. The U-Net was first proposed for biomedical image segmentation, but has found use in a lot of different applications. We make some modifications to it to make it suitable for our indoor imaging problem.

As explained in Section \ref{Sec_PS}, the DoI is divided into $N = n_x \times n_y$ grids. Using xRA, we obtain the initial contrast profiles in the form of a vector  $\overline{\raisedchi} = \big[(\operatorname{Re}({\overline{\raisedchi}_{\text{RI}}}))^T \  (	\operatorname{Im}({\overline{\raisedchi}_{\text{RI}}}))^T \big]^T$ where $\operatorname*{Re}(\overline{\chi}_{\text{RI}}), \operatorname*{Im}(\overline{\chi}_{\text{RI}}) \in \mathbb{R}^{N\times 1}$. These real and imaginary parts of the contrast vector are then converted to 2D images of dimensions $n_x \times n_y$ each representing the real and imaginary parts of contrast in every grid in the DoI ($\operatorname*{Re}(\overline{\overline{\chi}}_{\text{RI}}), \operatorname*{Im}(\overline{\overline{\chi}}_{\text{RI}}) \in \mathbb{R}^{n_x\times n_y}$). These are then given as input to the modified U-Net model as shown in Fig. \ref{unet}.

The U-net architecture we use is a symmetric encoder-decoder network with two input channels of dimensions $n_x \times n_y$, one each for $\operatorname*{Re}(\overline{\overline{\chi}}_{\text{RI}})$ and $ \operatorname*{Im}(\overline{\overline{\chi}}_{\text{RI}})$ and one output channel of dimensions $n_x \times n_y$ representing the final permittivity profile $\overline{\overline{\epsilon}}_R$. The encoder extracts low dimensional spatial features of the input and the decoder reconstructs $\overline{\overline{\epsilon}}_R$ from these encoded features. The input is passed through a series of layers that progressively downsample, until we get to a bottleneck layer, after which the process is reversed. Each encoder layer consists of two repeated applications of 3$\times$3 convolutional filters, each followed by batch normalization and a rectified linear unit (ReLU). This is followed by a 2$\times$2 max-pooling layer for downsampling. The number of channels doubles after each downsampling. Each layer of the decoder consists of a 2$\times$2 up-sampling layer followed by a 3$\times$3 convolution which halves the number of channels in each layer, eventually restoring the output to the same size as the input. The final layer consists of 1$\times$1 convolutions followed by ReLU activation to obtain $\overline{\overline{\epsilon}}_R$ in the output.

The U-Net also consists of shortcut connections (also known as skip connections) from the encoder to the decoder. Specifically, we add skip connections between each encoder layer \textit{i} and decoder layer \textit{n-i} where \textit{n} is the total number of layers in the encoder or decoder. Each skip connection concatenates all channels at the output of decoder layer \textit{n-i} with those at the output of encoder layer \textit{i} \cite{unetoriginal}.

The downsampling operation in the encoder leads to the loss of some information which could be vital for the decoder to reconstruct the output. The skip connections are highly effective in this case since they propagate the information lost by the encoder layers directly to the decoder. This is especially useful in cases when the input and output share similar features. However, for the inverse scattering use case dealing with scatterers of large size and high permittivity values (which none of the existing techniques consider), the input contains high amounts of distortion and noise, which might prove detrimental if directly concatenated to the decoder output. Also, encoder representations computed in the earlier layers of the network are supposed to be lower level representations. Through the skip connections, these are concatenated with the decoder representations which contain much higher level reconstructions since they are generated in the deeper layers of the network. Hence, there exists a possible semantic gap between the concatenated outputs. To mitigate these issues, we add convolutional layers into the skip connections as shown in Fig. \ref{unet}. Since the gap is the highest for the outermost skip connection and decreases as we go deeper, we reduce the number of convolutional layers as we go deeper as well. Specifically, for the skip connection between encoder layer \textit{i} and decoder layer \textit{n-i}, we add \textit{n-i} convolutional layers, each containing 3$\times$3 convolutional filters followed by batch normalization and ReLU activation. 

This modified U-Net architecture is suitable for the inverse scattering-based indoor imaging problem due to the following reasons: 

\begin{enumerate}
	\item \textbf{Capacity to model non-linearity:} For estimating $\overline{\overline{\epsilon}}_R$ given $\operatorname*{Re}(\overline{\overline{\chi}}_{\text{RI}})$ and $\operatorname*{Im}(\overline{\overline{\chi}}_{\text{RI}})$, we need to model the non-linear relationship mentioned in (\ref{Eq_newcontrast}). The modified U-Net is a type of Convolutional Neural Network (CNN), which is a sub-class of the more general feedforward neural networks. For such a network, the Universal Approximation Theorem stated in \cite{deeplearningbook, universalapproximation} means that a large enough network containing at least one hidden layer has the capacity to represent any continuous function, and this capacity increases as the depth of the network increases. Thus, the modified U-Net architecture shown in Fig. \ref{unet} has the capacity to model the non-linear relationship in (\ref{Eq_newcontrast}) to obtain $\overline{\overline{\epsilon}}_R$.
	\item \textbf{Translation Invariance:} Convolutional Neural Networks (CNNs) are known to be \textit{translation invariant} due to their use of convolutional and pooling layers \cite{deeplearningbook}. The convolutional filters are shared across multiple locations of the inputs $\operatorname*{Re}(\overline{\overline{\chi}}_{\text{RI}})$ and $\operatorname*{Im}(\overline{\overline{\chi}}_{\text{RI}})$ to account for the fact that they can have many statistical properties that are invariant to translation. Each pooling layer in the network also makes the layer representations slightly invariant to translation of the input in that layer. As we move to deeper layers, the receptive field of the pooling layer gets larger and larger, and we subsequently achieve translation invariance over the entire input. This property makes it possible for the modified U-Net model to reconstruct the accurate permittivity profile of the entire DoI regardless of where the scatterer lies in the DoI. 
	\item \textbf{Learning Priors:} The trained model acts as an implicit learned prior and embeds a detailed prior knowledge of structure in the input which it learns from training data. It does so on account of its encoder-decoder based architecture \cite{deepimageprior}. An important feature of this type of architecture is the bottleneck in the network caused by the downsampling of the input by the encoder layers followed by the upsampling of decoder layers. All information that passes through the network must pass through this bottleneck. Since this bottleneck restricts the amount of information that passes through, the network is forced to learn structure in data needed to reconstruct $\overline{\overline{\epsilon}}_R$ at the output.
\end{enumerate}

The network is trained using mean squared error loss and Adam optimizer with an initial learning rate of $3 \times 10^{-4}$ which is decayed exponentially with a rate of 0.8.

In the next section we present numerical results obtained using the modified U-net architecture shown in Fig. \ref{unet}, followed by experimental results in an indoor environment obtained from Wi-Fi RSSI data. 

\section{Simulation Results}
\label{Sec_NumericalResults}
This section describes simulation results to evaluate the performance of our proposed framework using simulated data. The setup used for the simulations is compatible with that used for experiments (experimental results are presented in the next section). To provide quantitative evaluation PSNR results are also provided for each reconstruction.

We compare our results with the output of the xRA inverse model (solved using Tikhonov regularization) itself. It is important to note that the xRA inverse model provides contrast profiles $\operatorname*{Re}(\chi_{\text{RI}})$ and $ \operatorname*{Im}(\chi_{\text{RI}})$, which are then given as input to the network to generate the $\epsilon_R$ profile of the DoI.

We do not provide comparisons with any other techniques as they also do not function well for the range of validity considered in this work.  To the best of our knowledge, there is only one other phaseless inverse scattering-based learning approach \cite{Xudongchen}, but it demonstrates results for scatterers with low permittivity values ($\epsilon_R < 4$) and size smaller than $\lambda_0$, which does not meet the validity range required for indoor imaging. It also uses non-linear inverse models to generate the initial contrast profile, which makes the process of generating training data highly time and compute-intensive. Finally, none of the existing techniques can be formulated for background subtraction, which is extremely crucial for accurate reconstruction in an actual indoor environment (where data can be highly distorted due to multipath reflections from ceiling, floor and other clutter). In contrast, our approach uses a linear inverse model, incorporates background subtraction and is shown to perform well for scatterers with extremely high permittivity values ($\epsilon_R > 50$) and large size which is well outside the range considered by other techniques.

\subsection{Problem Setup}
\label{Sec_NumericalSetup}

\begin{figure}[h]
	\centering
	\begin{subfigure}{0.21\textwidth}
		\centering
		\includegraphics[width=\textwidth]{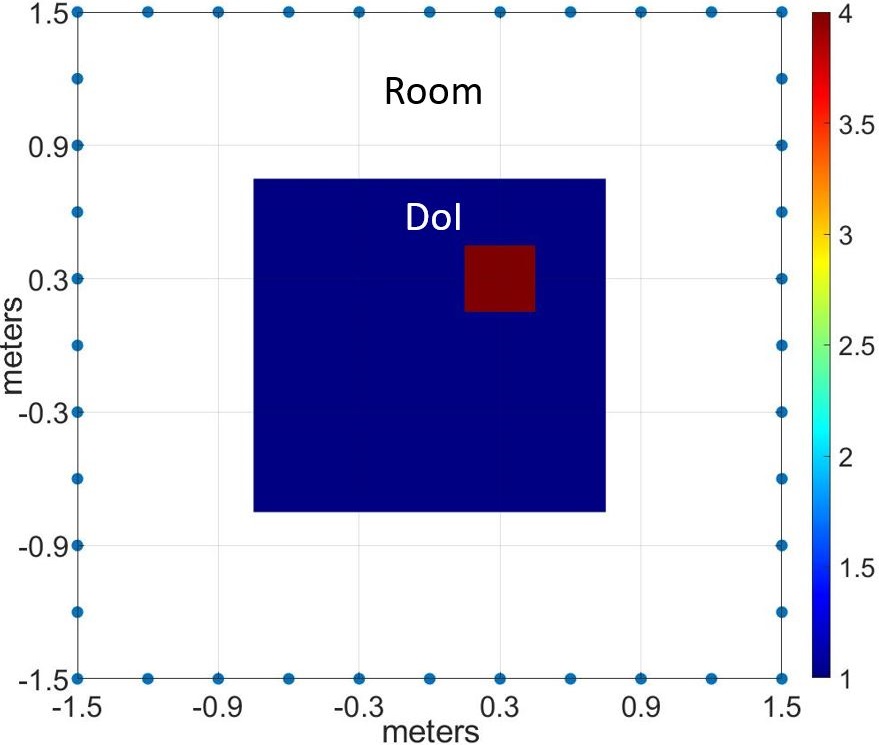}
		\subcaption{Numerical Setup}
	\end{subfigure}
	\begin{subfigure}{0.26\textwidth}
		\centering
		\includegraphics[width=\textwidth]{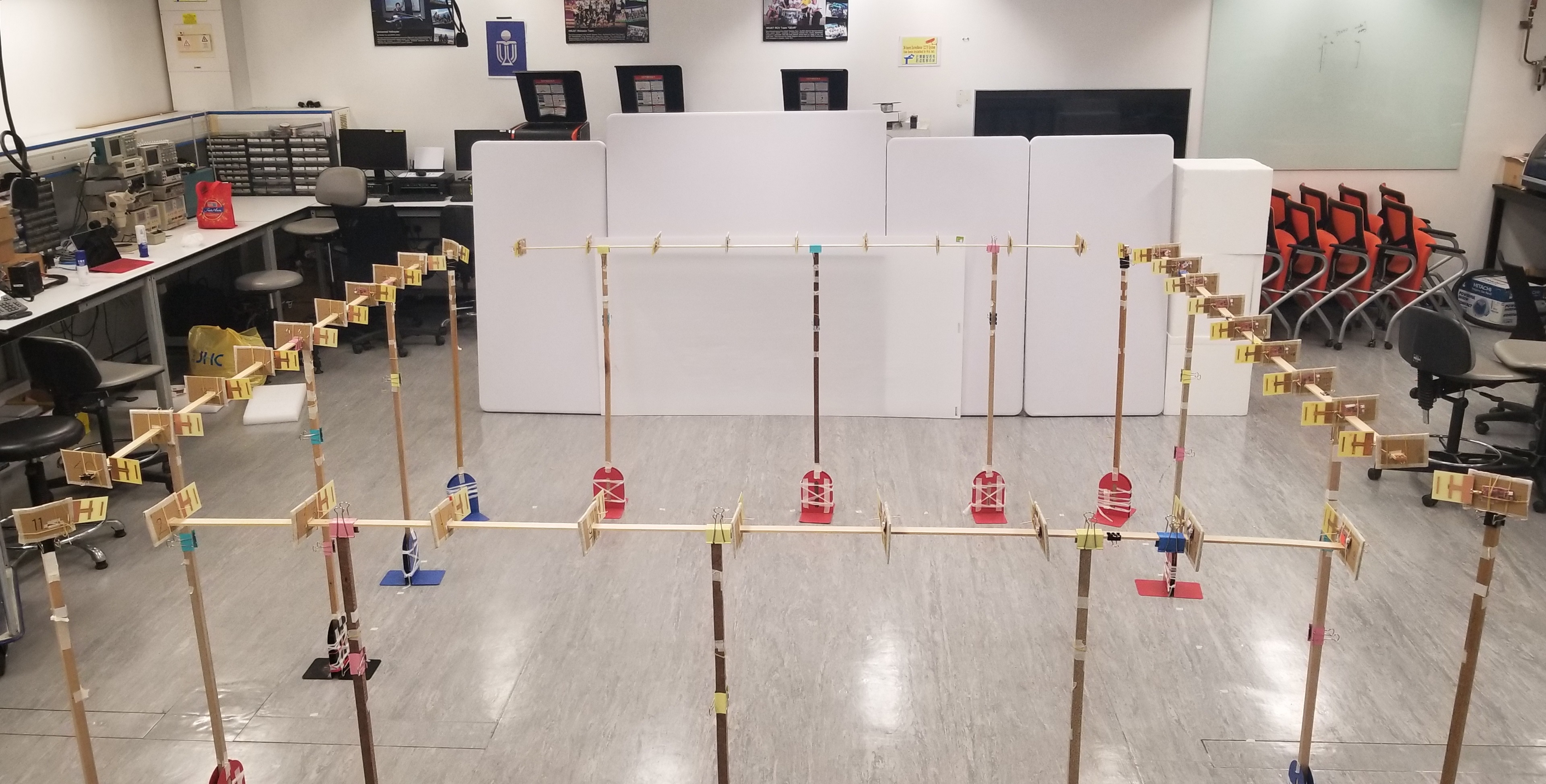}
		\subcaption{Experimental Setup}
	\end{subfigure}
	\caption{Data Acquisition Configurations}
	\label{problemsetup}
\end{figure}

For both simulations and experiments, we consider a $3\times3$ m$^2$ area containing a $1.5\times1.5$ m$^2$ DoI that needs to be imaged, both centered at the origin, as shown in Fig. \ref{problemsetup} (a). The DoI size is $12 \lambda_0 \times 12 \lambda_0$, where $\lambda_0$ = 0.125 m is the incident wavelength. There are 40 equidistant Wi-Fi transceivers ($M = 40$) operating at 2.4 GHz and placed at the boundary of the area to collect measurement data. When one transceiver transmits, the other 39 measure RSSI values. The total number of wireless links (or RSSI measurements) are $L= M(M-1)/2 = 780$ (note that we are not considering reciprocal links). We discretize the DoI into grids of 400$\times$400 (160000 grids) for the forward problem and 50$\times$50 (2500 grids) for the inverse problem.

To generate forward RSSI data for simulations and training, we use the Method of Moments approach detailed in \cite{chen2018computational}. This data is used to solve the inverse problem using the xRA inverse model detailed in Section \ref{Sec_RAf} to obtain the real and imaginary contrast profiles $\operatorname*{Re}(\overline{\overline{\chi}}_{\text{RI}}),  \operatorname*{Im}(\overline{\overline{\chi}}_{\text{RI}}) \in R^{50\times50}$. These contrast profiles are then given as input to the modified U-Net described in Section \ref{Sec_DLframework} which generates the final permittivity profile $\overline{\overline{\epsilon}}_R \in R^{50\times50}$.

\subsection{Training data}
\label{Sec_TrainingData}
For training the network, we generate a dataset of circular and square shaped lossless scatterers ($\epsilon_r = \epsilon_R$, $\epsilon_I = 0$) placed in the DoI. 
\begin{figure*}[!h]
	\centering
	\begin{subfigure}{\textwidth}
		\centering
		\includegraphics[scale=0.38]{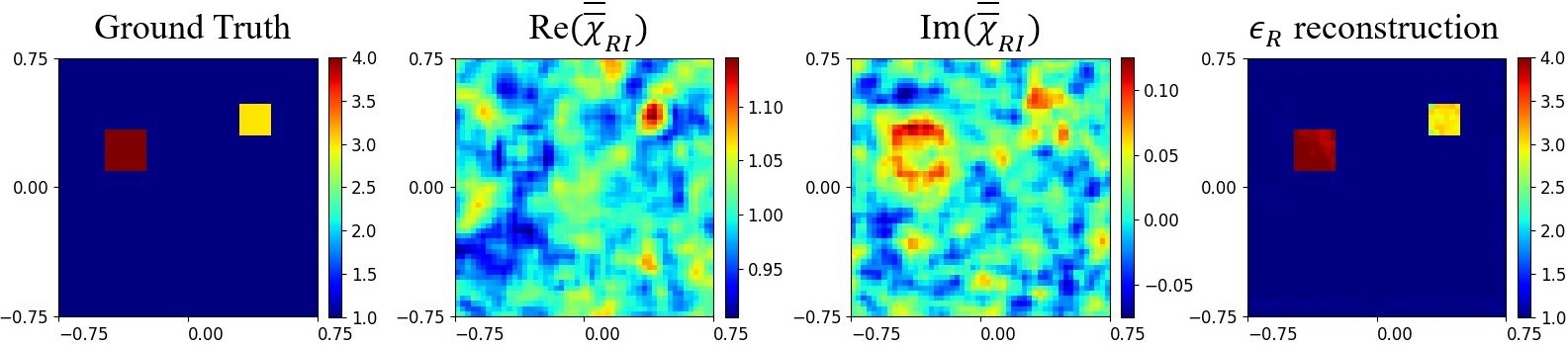}
		\subcaption{}
	\end{subfigure}
	\begin{subfigure}{\textwidth}
		\centering
		\includegraphics[scale=0.38]{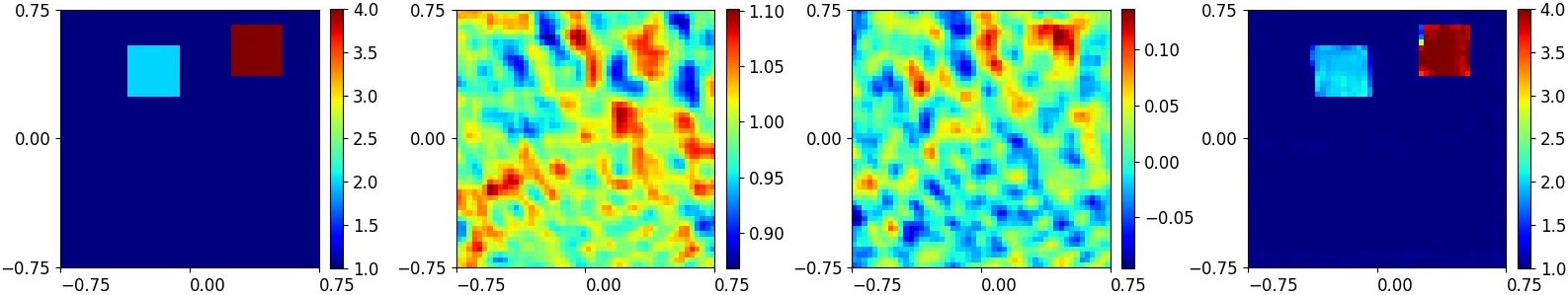}
		\subcaption{}
	\end{subfigure}
	\begin{subfigure}{\textwidth}
		\centering
		\includegraphics[scale=0.38]{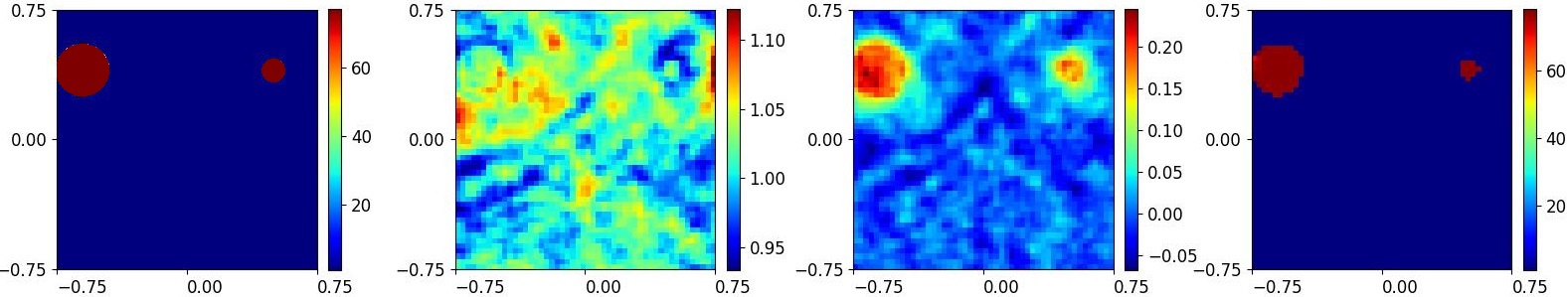}
		\subcaption{}
	\end{subfigure}
	\begin{subfigure}{\textwidth}
		\centering
		\includegraphics[scale=0.38]{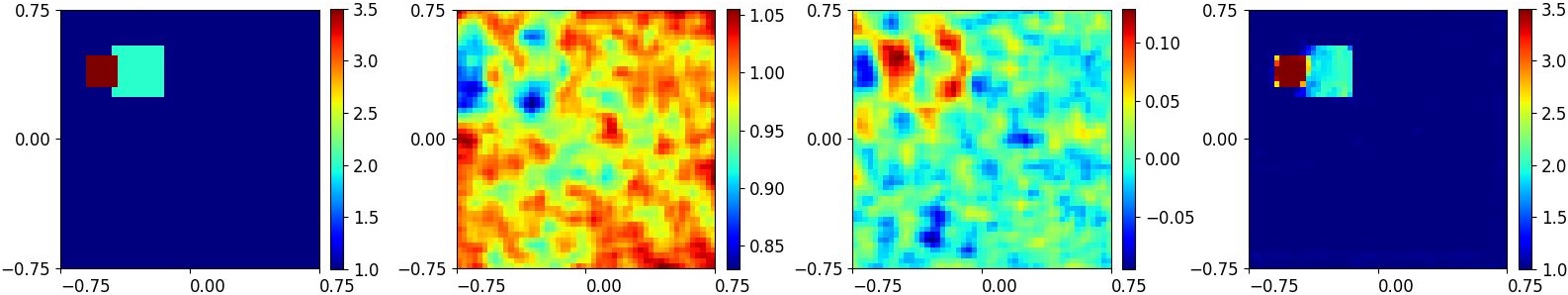}
		\subcaption{}
	\end{subfigure}
	\caption{Various reconstruction results where permittivity value is indicated by colour. First column shows the ground truth, the second and third columns show the real and imaginary parts of the extended Rytov approximation (xRA) reconstruction respectively, and the last column shows the final $\epsilon_R$  reconstruction using our physics assisted deep learning approach. The respective PSNR values of $\epsilon_R$ reconstructions are (a) 43.58 dB, (b) 31.72 dB, (c) 33.91 dB and (d) 35.19 dB.}
	\label{Results_ex1}
\end{figure*}

\begin{figure*}[!h]
	\centering
	\begin{subfigure}{\textwidth}
		\centering
		\includegraphics[scale=0.385]{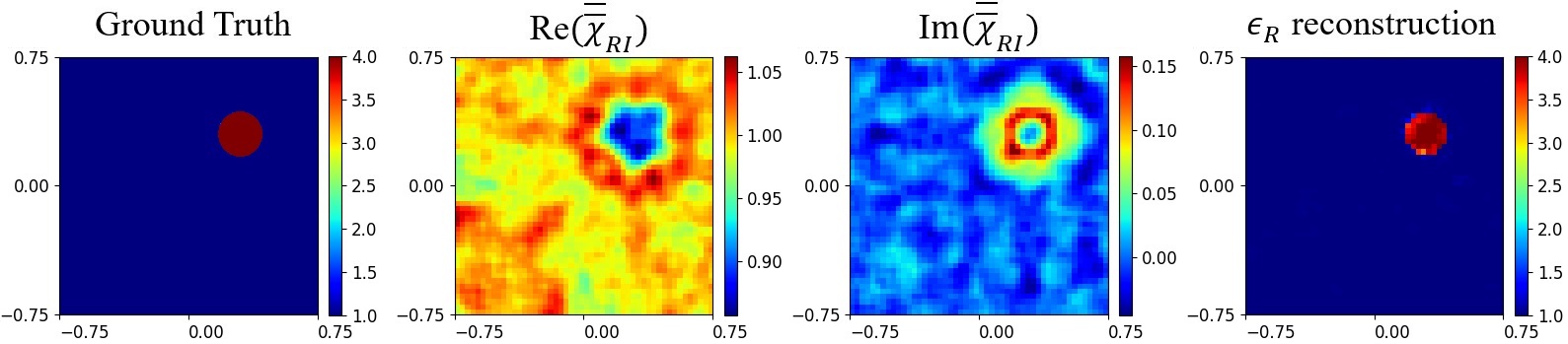}
		\subcaption{}
	\end{subfigure}
	\begin{subfigure}{\textwidth}
		\centering
		\includegraphics[scale=0.48]{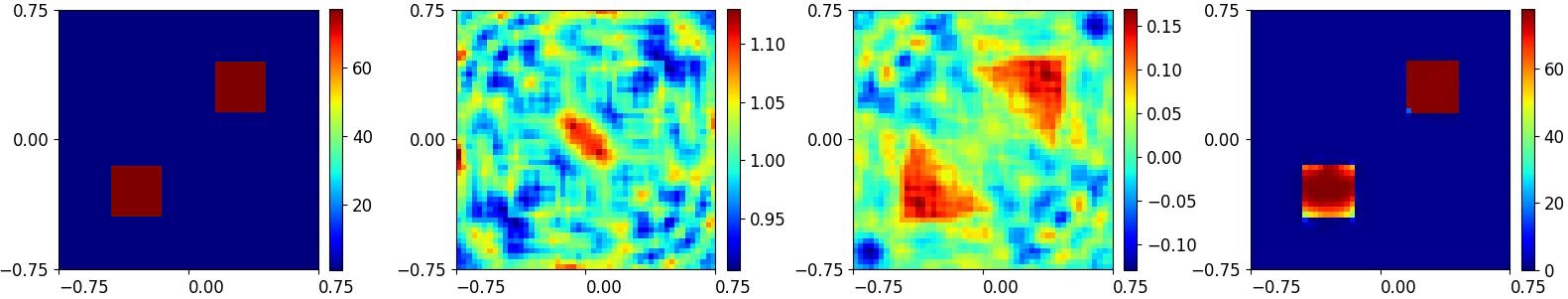}
		\subcaption{}
	\end{subfigure}
	\begin{subfigure}{\textwidth}
		\centering
		\includegraphics[scale=0.48]{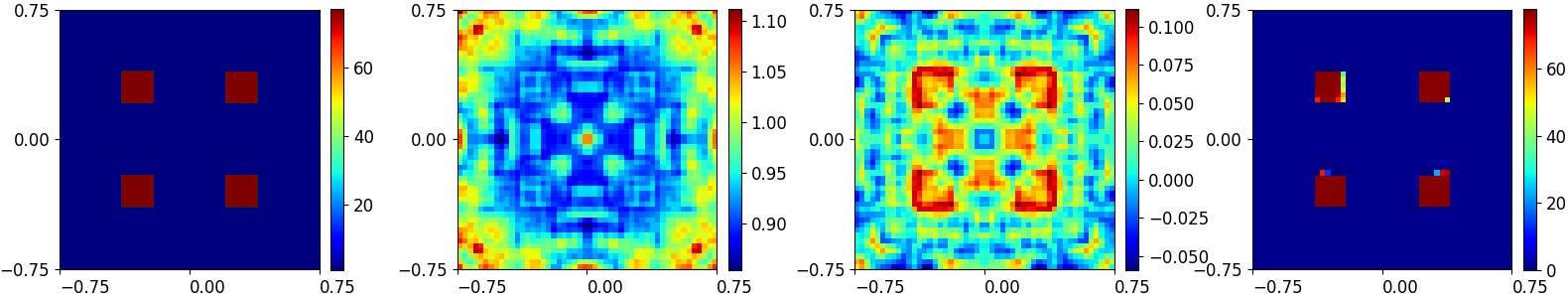}
		\subcaption{}
	\end{subfigure}
	\begin{subfigure}{\textwidth}
		\centering
		\includegraphics[scale=0.48]{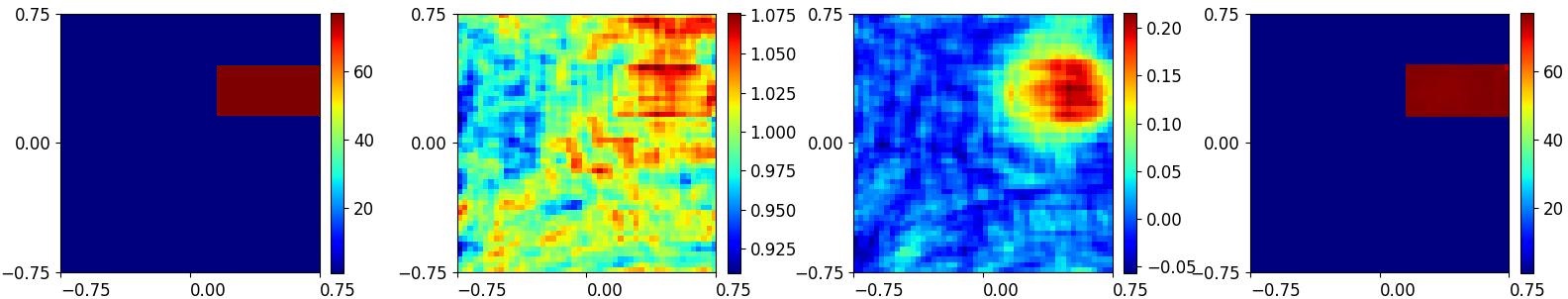}
		\subcaption{}
	\end{subfigure}
	\begin{subfigure}{\textwidth}
		\centering
		\includegraphics[scale=0.48]{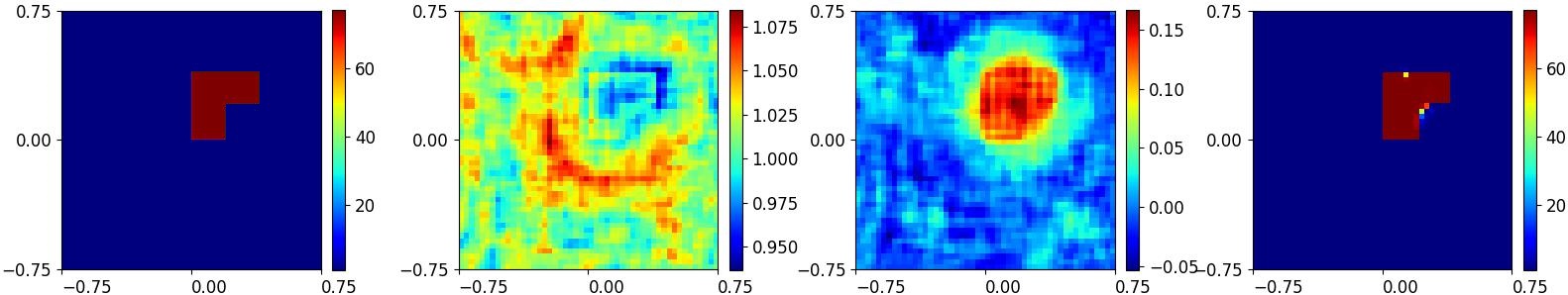}
		\subcaption{}
	\end{subfigure}
	\caption{Various reconstruction results where permittivity value is indicated by colour. First column shows the ground truth, the second and third columns show the real and imaginary parts of the xRA reconstruction respectively, and the last column shows the final reconstruction using our physics assisted deep learning approach. The respective PSNR values of $\epsilon_R$ reconstructions are (a) 39.58 dB, (b) 30.23 dB, (c) 28.11 dB, (d) 52.13 dB and (e) 30.66 dB.}
	\label{Results_ex2}
\end{figure*}

A single scatterer is chosen to be a square or a circle with an equal probability, and lies completely in the DoI. Due to the \textit{translational invariance} property of CNNs mentioned in Section \ref{Sec_DLframework}, we place scatterers only in the top-half of the DoI, which greatly reduces the amount of data needed for the modified U-Net to learn the non-linear relation in (\ref{Eq_newcontrast}). Thus, assuming that the origin lies at the center of the DoI, the $x$ and $y$ co-ordinates of the scatterer centers are sampled uniformly from $c_x=(-0.6, 0.6)$ and $c_y=(0.15, 0.6)$ respectively. 

To select the real component of permittivity values $\epsilon_R$ for the randomly generated scatterers, we use prior information about the typical range of permittivity values for various objects found in the indoor region. For example, most objects (such as wood, concrete walls, furniture and plastic) have permittivity values in the range $1 < \epsilon_R< 5$ at 2.4 GHz, whereas any object containing water (like the human body) can have permittivity values in the range $50<\epsilon_R<77$. Therefore, we sample permittivity values uniformly from these two intervals (i.e. $1<\epsilon_R<5$ and $50<\epsilon_R<77$) at an interval of $0.2$.

The size of the scatterers (length of diameter for circular scatterers and length of side for square scatterers) is also sampled uniformly from the set $S_{L} \in \{0.5\lambda_0, \lambda_0, 1.5\lambda_0, 2\lambda_0, 2.5\lambda_0\}$. Therefore, if we consider a scatterer of size 2.5 $\lambda_0$ with $\epsilon_R = 77$, its electrical size is $\approx 20\lambda$ where $\lambda = \lambda_0/\sqrt{77}$.

Since we consider permittivity values as high as $\epsilon_R = 77$, multiple scattering effects among scatterers are expected to be significant. The network cannot predict these effects if it is only trained on samples containing single scatterers. Therefore, we train the network using samples containing two scatterers each. As shown in the following sections, this generalizes well to settings where the test data contains only one scatterer or more than two scatterers.

We generate 5000 samples for training, each of which contains two scatterers with the aforementioned specifications. We use 500 of these samples as test data and divide the rest into training and validation datasets, using 4000 samples for training and 500 samples for validation.

\subsection{Reconstruction Results}
\label{Sec_NumericalTests}

We use the modified UNet model as a part of our framework since it gives better reconstructions than UNet for our inverse scattering use case ({as shown in the next subsection}).

The reconstructed profiles of different tests are presented in Figs. \ref{Results_ex1} and \ref{Results_ex2} where the PSNR results are listed in the figure captions. Fig. \ref{Results_ex1} contains samples from the test dataset and Fig. \ref{Results_ex2} contains samples which belong to neither training nor test datasets and can provide generalization tests for the proposed framework. Most of the scatterers in these tests have a size comparable to or larger than $\lambda_0$ and the largest size can go up to 5$\lambda_0$. In contrast, all of the existing techniques start to break down when the scatterer size approaches $\lambda_0$. 

We can see in all the examples in Figs. \ref{Results_ex1} and \ref{Results_ex2} that our proposed framework provides accurate shape reconstructions for all scatterers, even when the xRA reconstructions given as input to the modified U-Net model do not show any distinctive shapes in most cases. This is because the trained model learns prior information about scatterer shapes from training data as explained in Section \ref{Sec_DLframework}.

We can also see that the predicted permittivity values of all scatterers are very close to their actual values. Fig. \ref{Results_ex1} (a), (b) and (d) show a clear distinction in the output for scatterers with different permittivity values. The result is especially interesting for Fig. \ref{Results_ex1} (c), in which both scatterers have $\epsilon_R = 77$, which is significantly higher than the validity range of any of the existing techniques. At such high permittivity values, multiple scattering effects both within scatterers and between two scatterers are very prominent. The accurate reconstruction of $\epsilon_R$ in such cases can be directly attributed to the model described in Section \ref{Sec_RAf} since the reconstructions of both $\operatorname*{Re}(\chi_{\text{RI}})$ and $ \operatorname*{Im}(\chi_{\text{RI}})$ obtained contain information about $\epsilon_R$. As explained in Section \ref{Sec_DLframework}, the modified U-net then learns the non-linear relation in (\ref{Eq_newcontrast}), which leads to highly accurate reconstruction of the $\epsilon_R$ profile with precise permittivity values.

Fig. \ref{Results_ex2} (a) and (c) show that the network trained on samples containing two scatterers can easily generalize to one scatterer or even more than two scatterers. We can also see the translational invariance property of CNNs at play in Fig. \ref{Results_ex2} (b) and (c) where the scatterers present in the lower half of the DoI are also reconstructed accurately (in spite of the network being trained only on samples containing scatterers in the upper half of the DoI). Fig. \ref{Results_ex2} (d) shows a scatterer with physical length 5$\lambda_0$ and $\epsilon_R$ = 77. Thus, we even get accurate shape and permittivity value reconstruction for a scatterer with electrical length $5\lambda_0 \times \sqrt{77}$ $\approx$ 40$\lambda$.

For each of these tests, the generation of the xRA reconstructions takes an average of 1.4 seconds and the $\epsilon_R$ reconstruction using the modified UNet takes an average of 0.06 seconds on a system with Intel(R) Core(TM) i7-6700K 3.4GHz CPU with 16 GB RAM. This can be further accelerated using GPU based systems.

To summarize, our simulation results show accurate reconstruction of scatterers with very high permittivity values and large size, which has not been demonstrated earlier. Reconstruction PSNR is greater than 30 dB for all examples. The proposed framework also generalizes to shapes and locations of scatterers not present in the training data.

\subsection{Comparison between UNet and Modified UNet}
\label{compare_unet}
\begin{figure}[!h]
	\centering
	\begin{subfigure}{0.24\textwidth}
		\centering
		\includegraphics[scale=0.3]{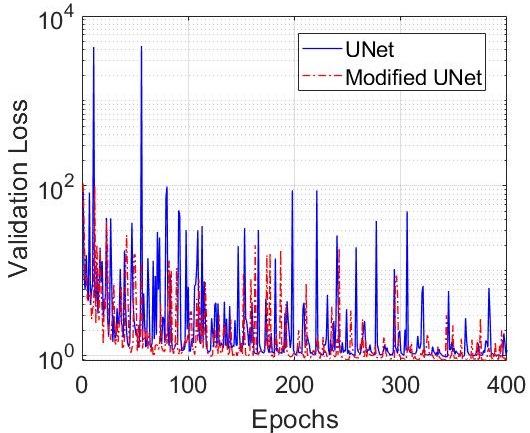}
		\subcaption{Original curves}
	\end{subfigure}
	\begin{subfigure}{0.24\textwidth}
		\centering
		\includegraphics[scale=0.3]{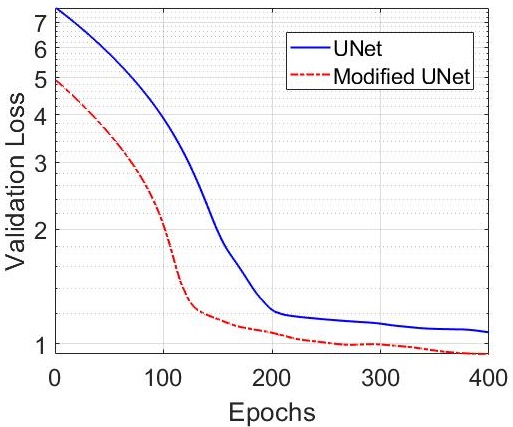}
		\subcaption{Smoothed curves}
	\end{subfigure}
	\caption{Validation loss for UNet and modified UNet architectures plotted on log scale.}
	\label{Results_dlcurves}
\end{figure}

Fig. \ref{Results_dlcurves}(a) shows the validation loss for the UNet and modified UNet models described in Section \ref{Sec_DLframework} when both of them are trained on the training data described in Section \ref{Sec_TrainingData} for a total of 400 epochs. Fig. \ref{Results_dlcurves}(b) shows the same graph smoothed using moving average for better interpretation (using the \textit{smooth()} function in MATLAB with a window size of 50). We can see from Fig. \ref{Results_dlcurves}(b) that the modified UNet model achieves a lower validation loss as compared to the original UNet model.

\begin{figure}[!h]
	\centering
	\begin{subfigure}{0.5\textwidth}
		\centering
		\includegraphics[scale=0.20]{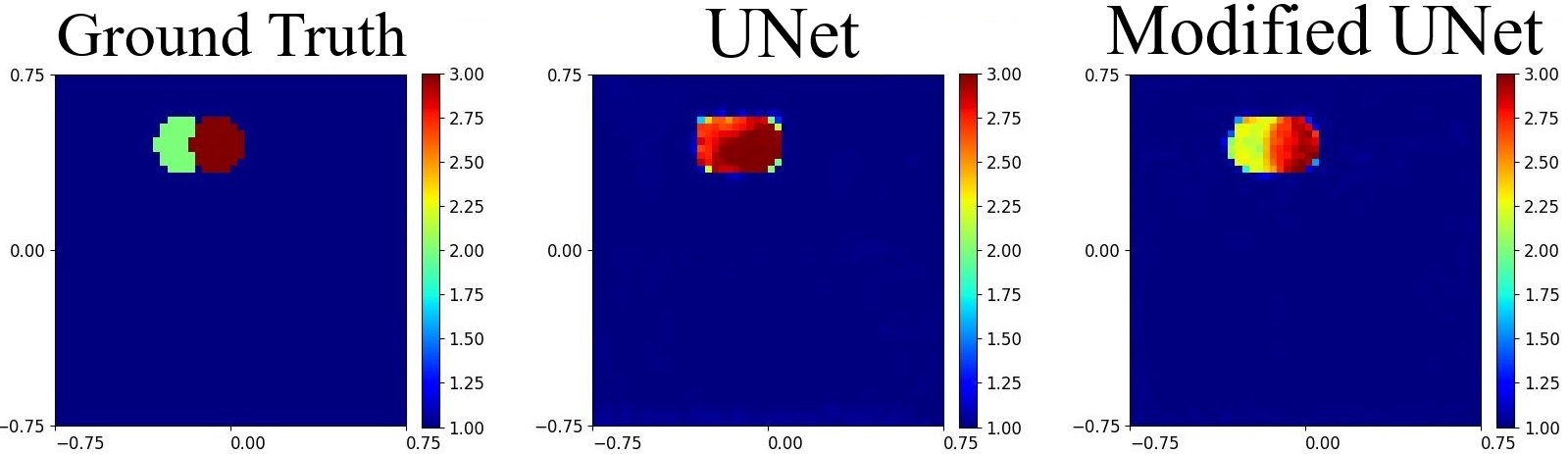}
		\subcaption{}
	\end{subfigure}
	\begin{subfigure}{0.5\textwidth}
		\centering
		\includegraphics[scale=0.20]{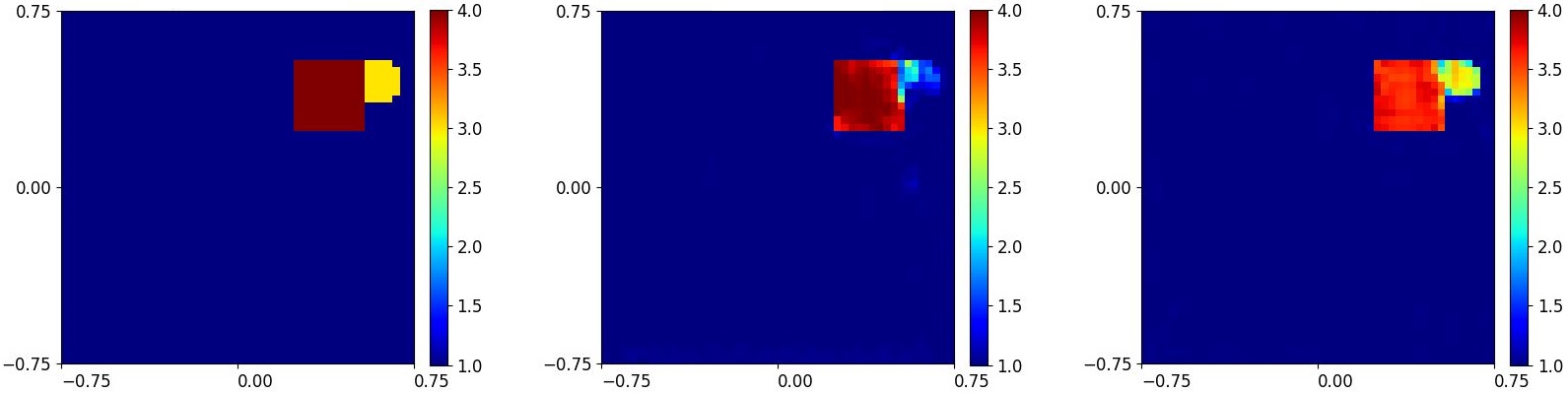}
		\subcaption{}
	\end{subfigure}
	\begin{subfigure}{0.5\textwidth}
		\centering
		\includegraphics[scale=0.20]{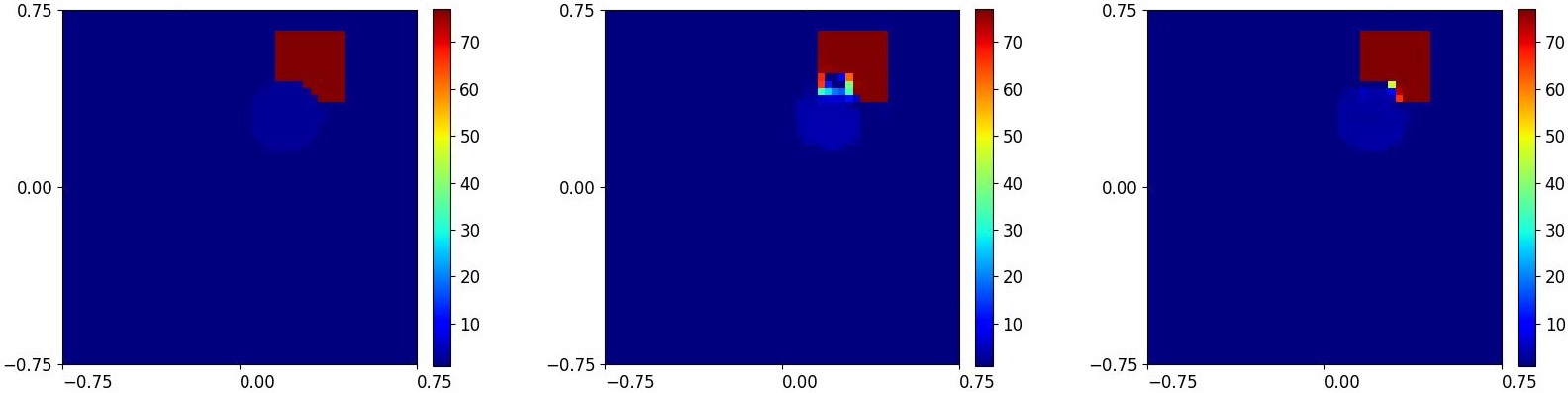}
		\subcaption{}
	\end{subfigure}
	\begin{subfigure}{0.5\textwidth}
		\centering
		\includegraphics[scale=0.20]{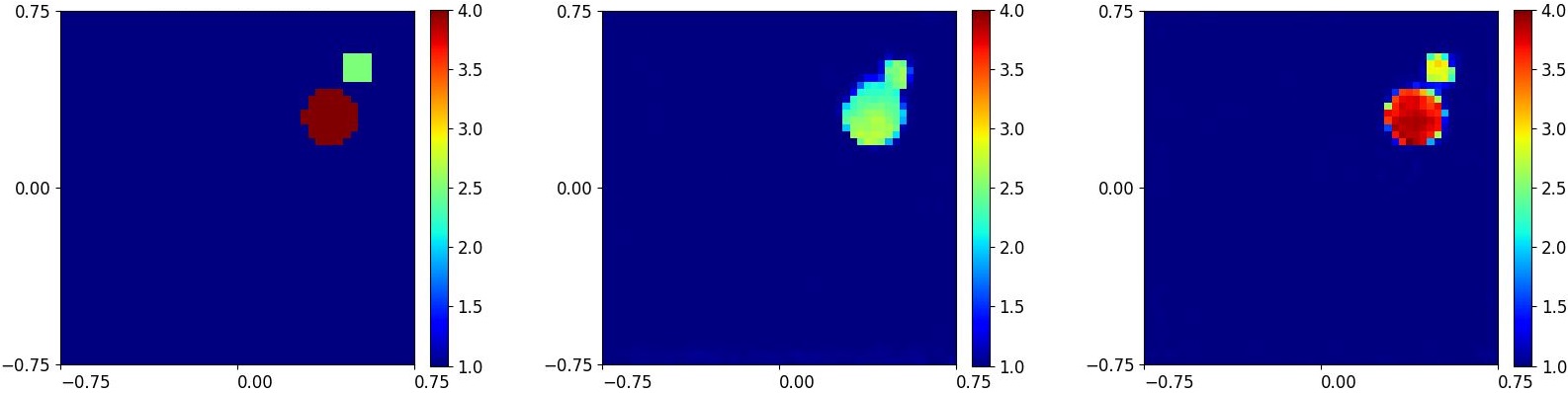}
		\subcaption{}
	\end{subfigure}
	\caption{Reconstructions for $\epsilon_R$ using the original UNet architecture and the modified UNet architecture used in this work. First column represents the ground truth data and the second and third columns show $\epsilon_R$ reconstructions obtained using UNet and modified UNet models respectively.}
	\label{Results_dl_compare}
\end{figure}

The difference between the final loss values of the two models might not seem highly significant at first glance. However, it is important to note that since the scatterers are much smaller than the DoI and we are dealing with the use case of imaging change in an indoor region, the model outputs are highly sparse. So the error due to inaccurate reconstructions of shape and permittivity values only occurs in a small region of the DoI where scatterers are placed. Due to this, even a small decrease in the validation loss translates to more accurate reconstruction of the shape and permittivity values. This is evident from the examples shown in Fig. \ref{Results_dl_compare} where the modified UNet model provides better shape and permittivity reconstructions. This is also supported by the PSNR values of these reconstructions listed in Table \ref{Table_dlcompare}.

\begin{table}[!h]
	\begin{center}
		\begin{tabular}{|c|c|c|}
			\hline
			Example & UNet & Modified UNet \\ [0.5ex] 
			\hline
			Fig. \ref{Results_dl_compare}(a) & 26.82 & 26.90 \\ [1ex] 
			\hline
			Fig. \ref{Results_dl_compare}(b) & 26.19 & 29.75 \\ [1ex] 
			\hline
			Fig. \ref{Results_dl_compare}(c) & 27.61 & 34.51 \\ [1ex] 
			\hline
			Fig. \ref{Results_dl_compare}(d) & 24.25 & 32.28 \\ [1ex] 
			\hline
		\end{tabular}
	\end{center}
	\caption{PSNR values for reconstructions shown in Fig. \ref{Results_dl_compare}.}
	\label{Table_dlcompare}
\end{table}

Also, we can see in Fig. \ref{Results_dlcurves}(b) that for any epoch, the smoothed validation loss of the modified UNet is lower than of UNet which shows that the variance of validation loss when using the modified UNet is lower. This might be due to the fact that instead of concatenating the input and encoder outputs which contain high distortion and noise directly to decoder layer outputs, in the modified UNet the encoder outputs go through additional convolutional layers in the skip connections, thus reducing noise in them and leading to better reconstructions. However, a more thorough analysis of improvement obtained by using the modified UNet model needs to be performed as part of future work using a series of controlled experiments. Other architectural changes to make the model even more suitable for the inverse scattering problem can also be explored.

\section{Experimental Results}
\label{Sec_ExperimentalResults}

\subsection{Experimental Setup}
\label{Sec_ExperimentalSetup}
Fig. \ref{problemsetup} (b) shows the experimental setup we use for indoor imaging. Wi-Fi transceiver nodes can be seen at the edge of the 3$\times$3 m$^{2}$ area. The environment is approximated as a two-dimensional (2D) electromagnetic problem where the DoI to be imaged is considered to be a 2D planar cross-section parallel to the floor at a height of 1.2 m. This 2D approximation is valid for the assumption that there is negligible scattering from floor, ceiling and other clutter so that the scattering is mainly from objects kept in the DoI cross-section. We achieve this by using directive antennas at the nodes and also using the background subtraction framework of xRA as explained in Section \ref{Sec_TBS}. Similar to the setup used for simulations detailed in Section \ref{Sec_NumericalSetup}, the experimental setup also uses $M=40$ Wi-Fi transceiver nodes, so there are a total of $L=M(M-1)/2=780$ unique measurement links. Obtaining one set of $L$ measurements takes around 10 seconds. We acquire $6$ such sets of measurements over a span of one minute and average them to use as final measurements. This makes our framework robust against missed transmissions or receptions at the transceivers. Each transceiver node consists of a SparkFun ESP32 Thing board \cite{ESP3200} consisting of an integrated 802.11 b/g/n Wi-Fi transceiver operating at 2.4 GHz and is placed at a height of 1.2 m from the floor using wooden stands. To minimize scattering from the floor, ceiling and clutter outside the DoI, the inbuilt omnidirectional antennas of the SparkFun ESP32 boards are replaced by directive Yagi antennas of 6.6 dBi. The 1.5$\times$1.5 m$^{2}$ DoI is discretized into grids of 50$\times$50 (2500 grids) for the inverse problem.

\begin{figure*}
	\centering
	\begin{subfigure}{\textwidth}
		\centering
		\includegraphics[scale=0.032]{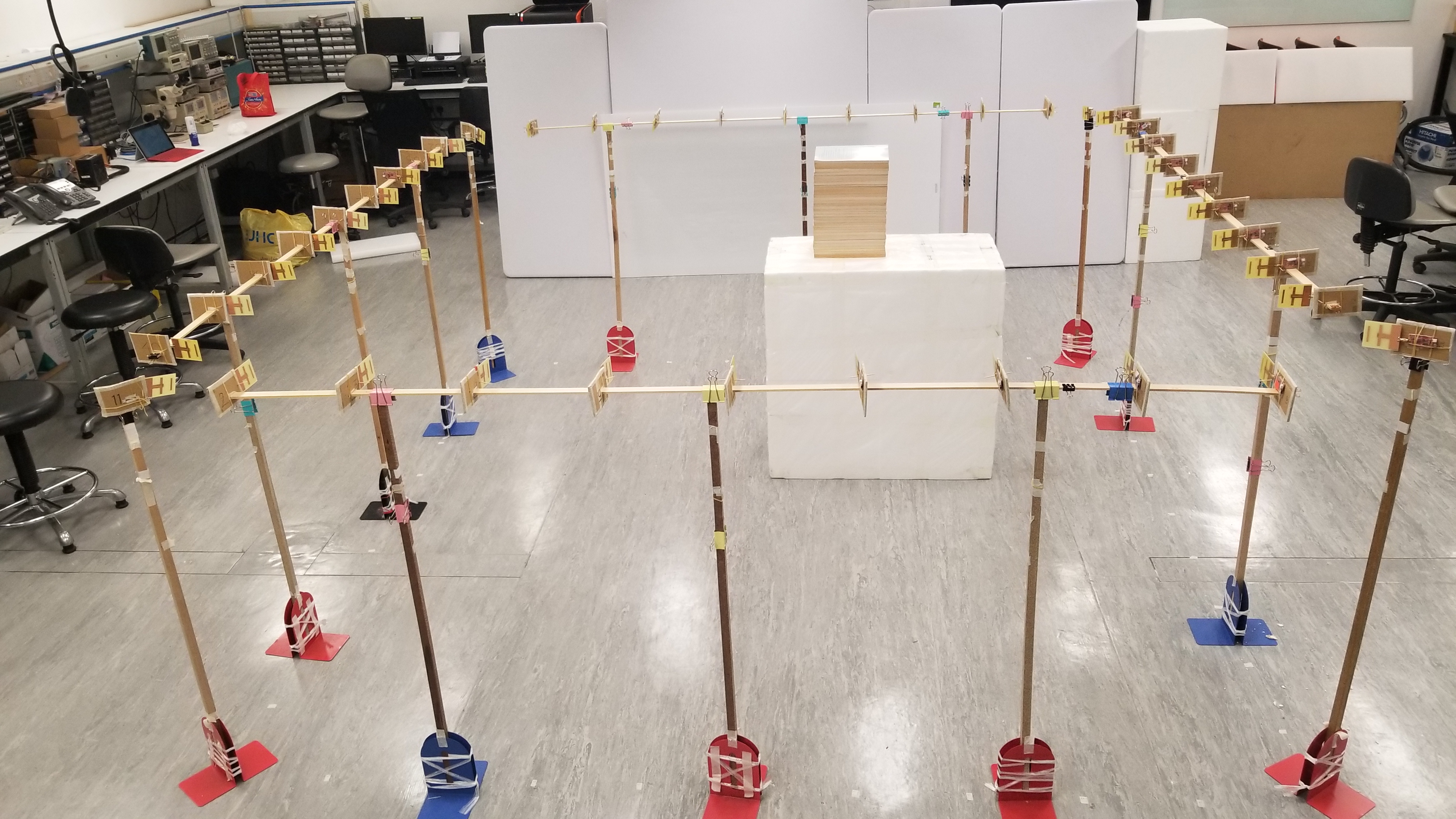}
		\includegraphics[scale=0.19]{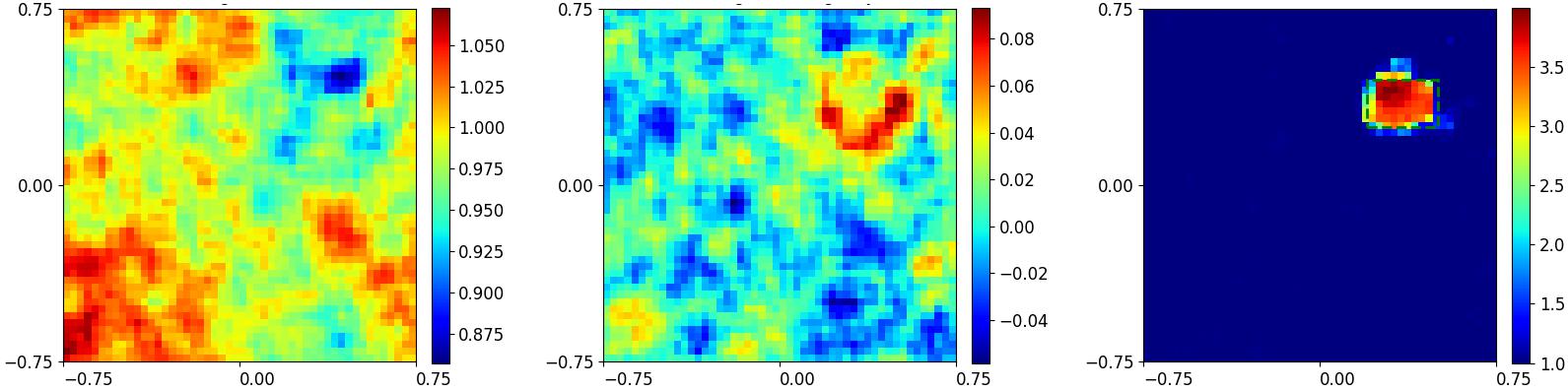}
		\subcaption{Single stack}
	\end{subfigure}
	\begin{subfigure}{\textwidth}
		\centering
		\includegraphics[scale=0.032]{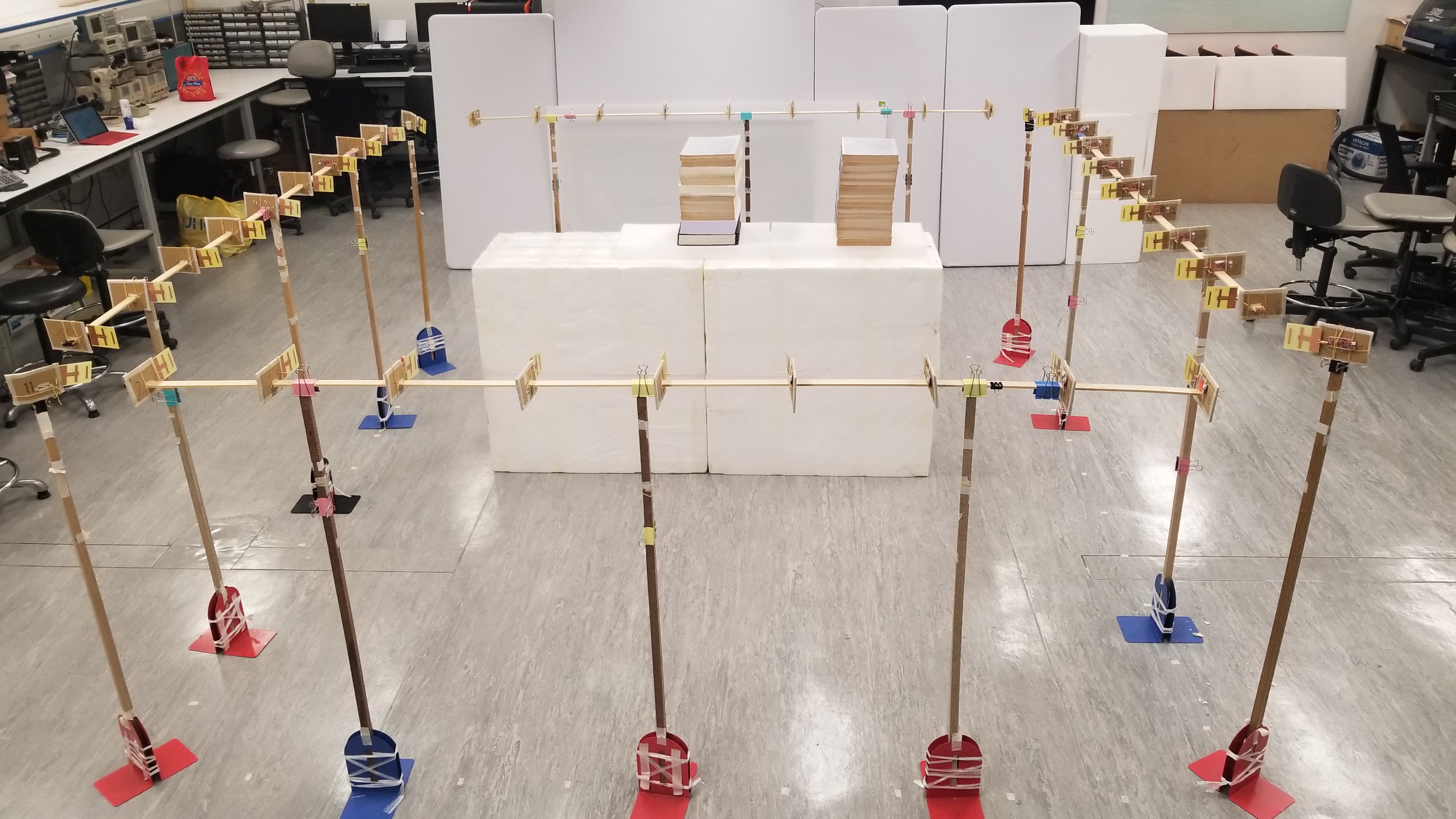}
		\includegraphics[scale=0.19]{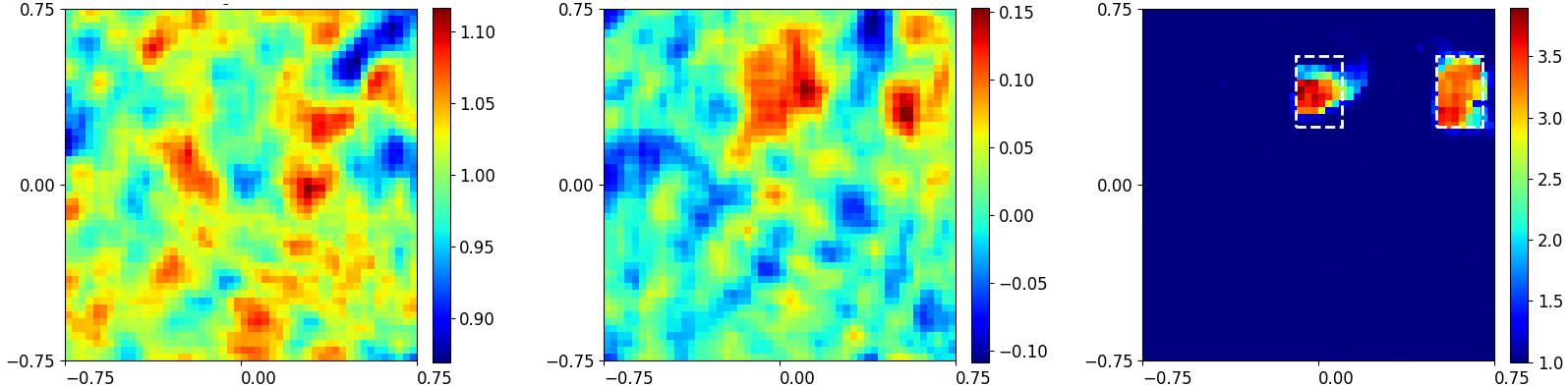}
		\subcaption{Two stacks placed at a distance}
	\end{subfigure}
	\begin{subfigure}{\textwidth}
		\centering
		\includegraphics[scale=0.032]{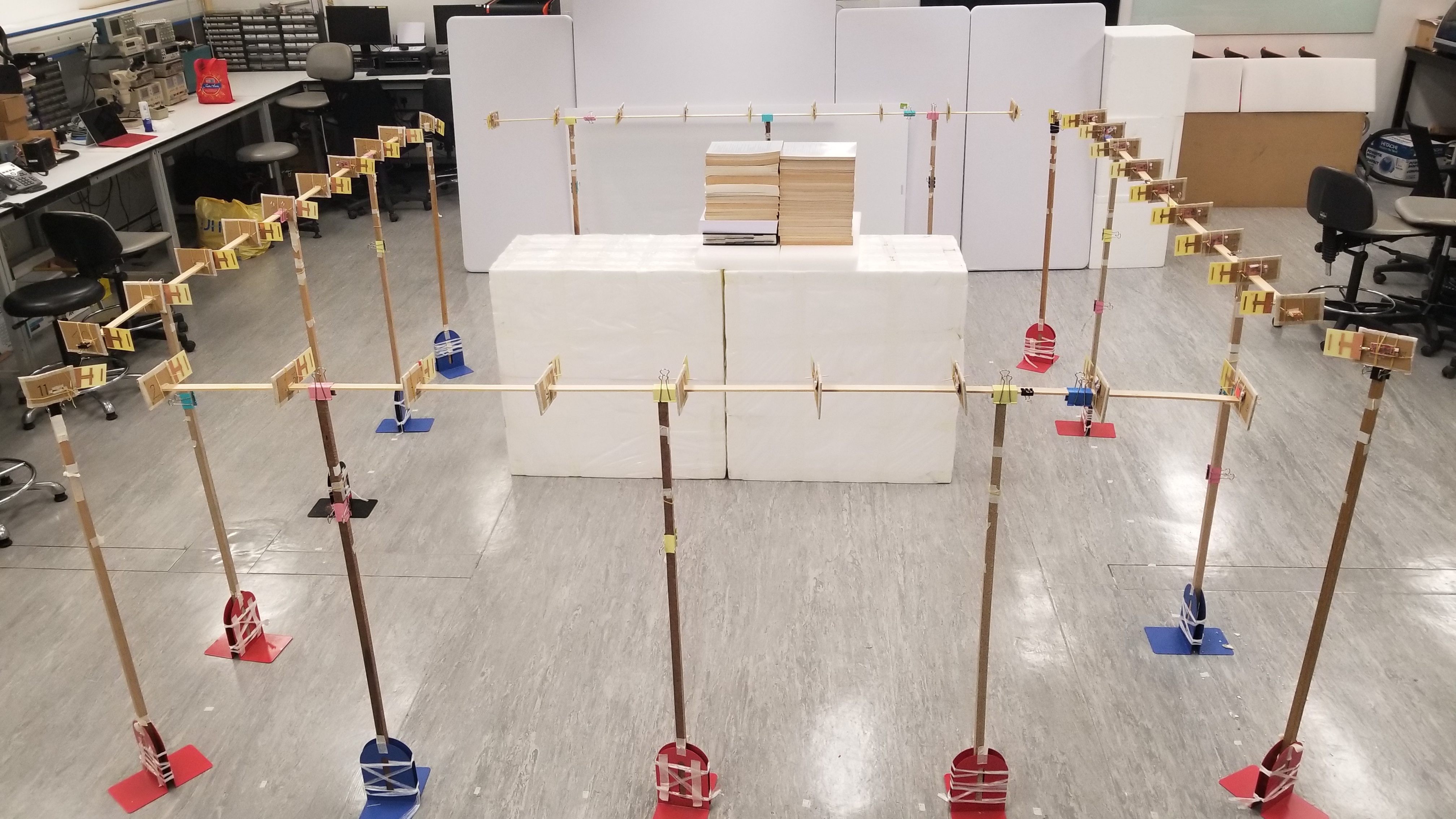}
		\includegraphics[scale=0.19]{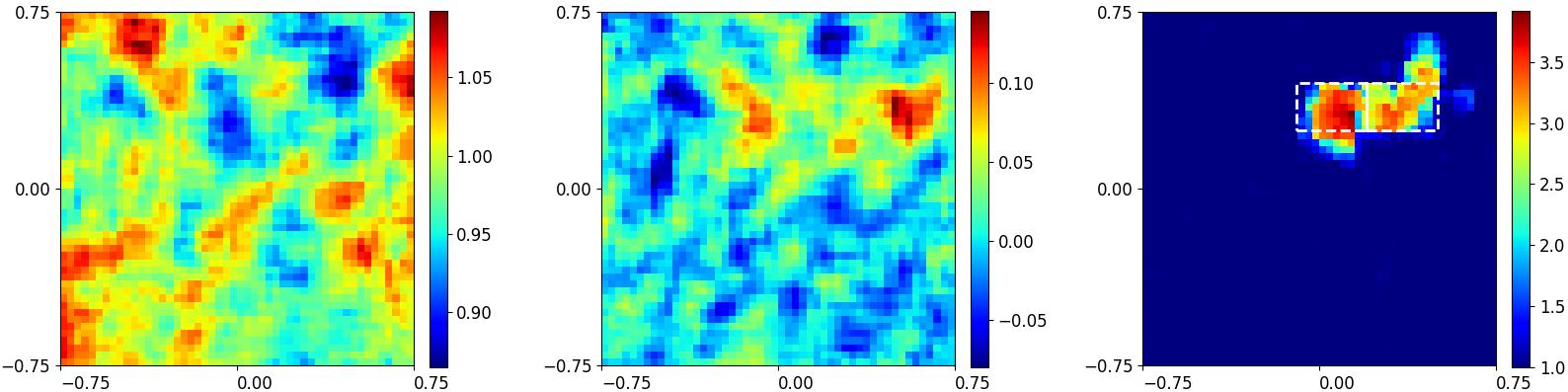}
		\subcaption{Two stacks joined together}
	\end{subfigure}
	\caption{Reconstruction results for one or more stacks of books ($\epsilon_r = 3.4 + j 0.25$). First column shows the experimental ground truth, the second and third columns show the real and imaginary parts of the xRA reconstruction respectively, and the last column shows the final $\epsilon_R$ reconstruction using our physics assisted deep learning approach. The respective PSNR values of $\epsilon_R$ reconstructions are (a) 23.50 dB, (b) 21.03 dB and (c) 19.36 dB.}
	\label{Results_books}
\end{figure*}

\begin{figure*}
	\centering
	\begin{subfigure}{\textwidth}
		\centering
		\includegraphics[scale=0.033]{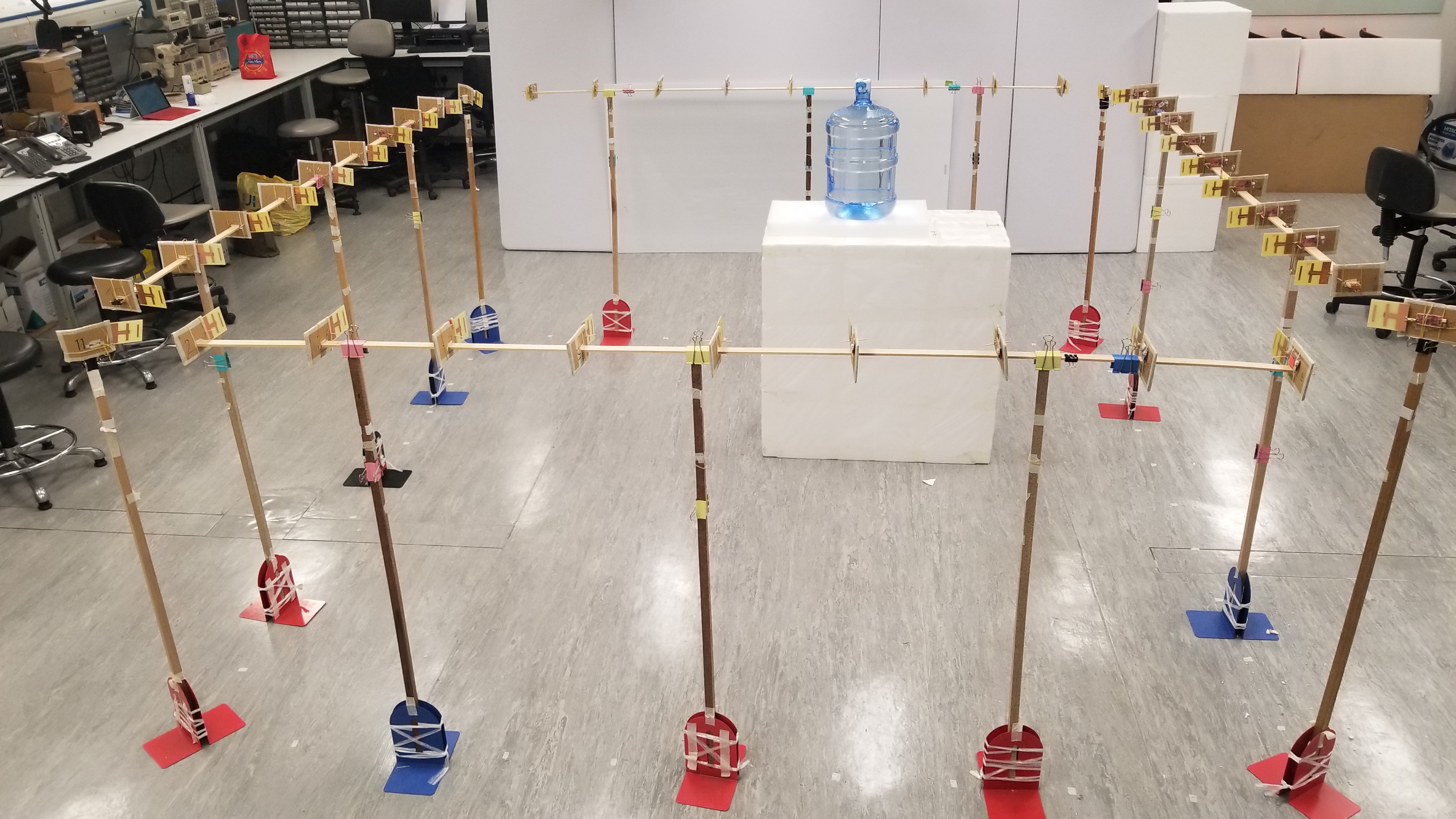}
		\includegraphics[scale=0.19]{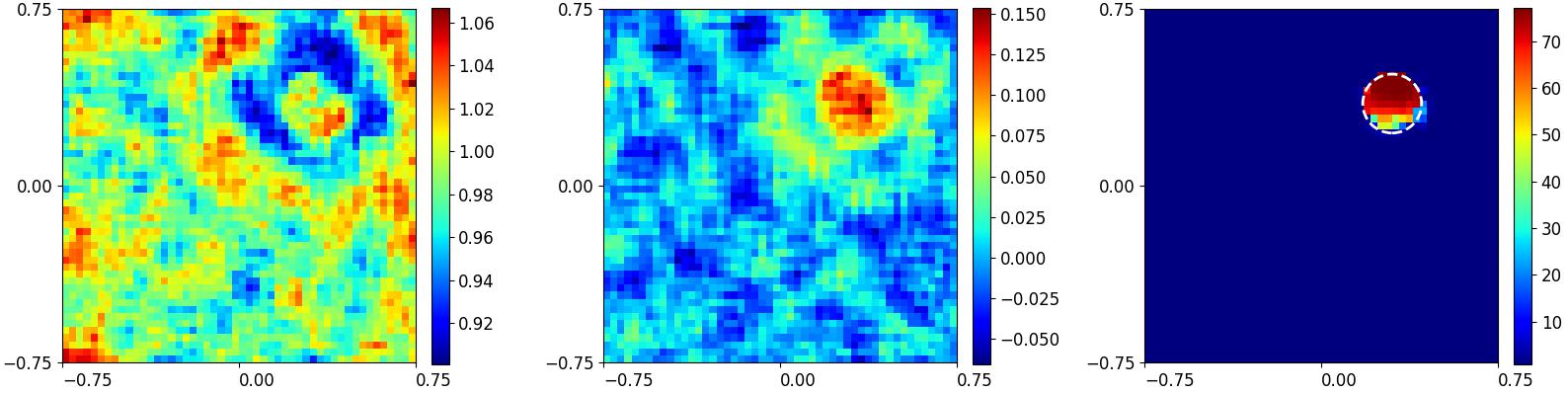}
		\subcaption{Single container}
	\end{subfigure}
	\begin{subfigure}{\textwidth}
		\centering
		\includegraphics[scale=0.033]{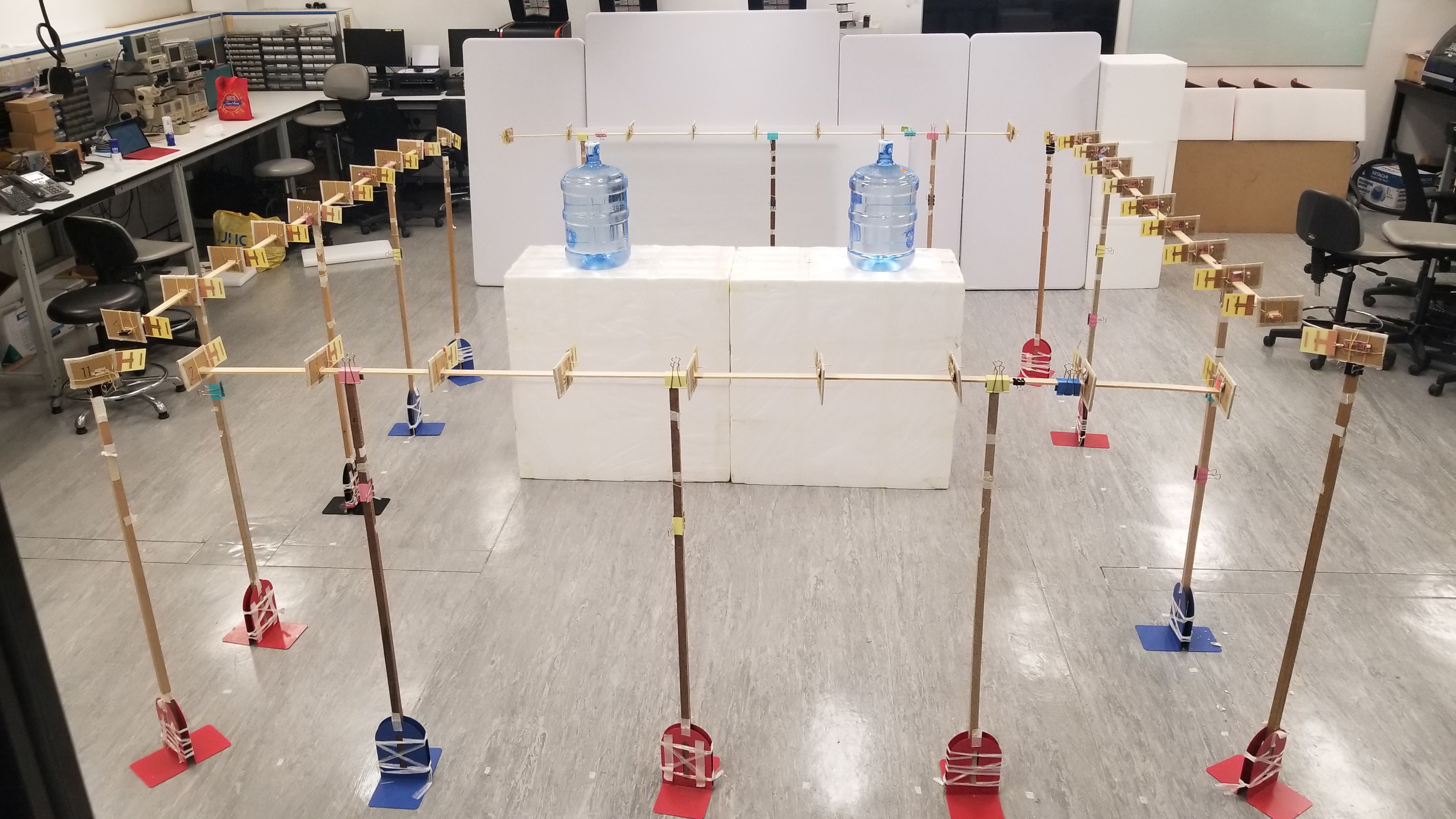}
		\includegraphics[scale=0.19]{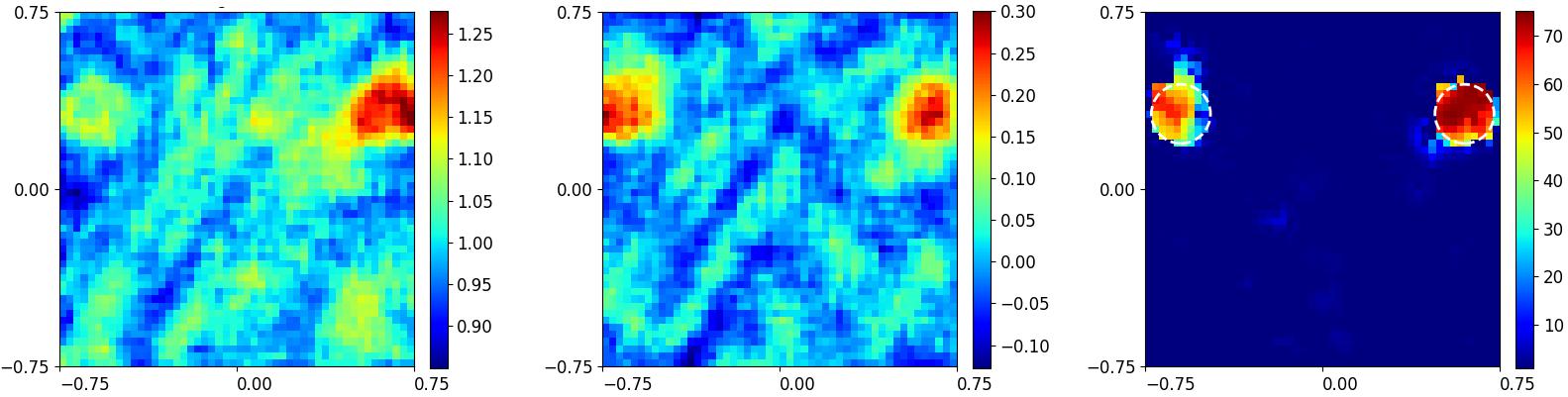}
		\subcaption{Two containers placed at a distance}
	\end{subfigure}
	\begin{subfigure}{\textwidth}
		\centering
		\includegraphics[scale=0.033]{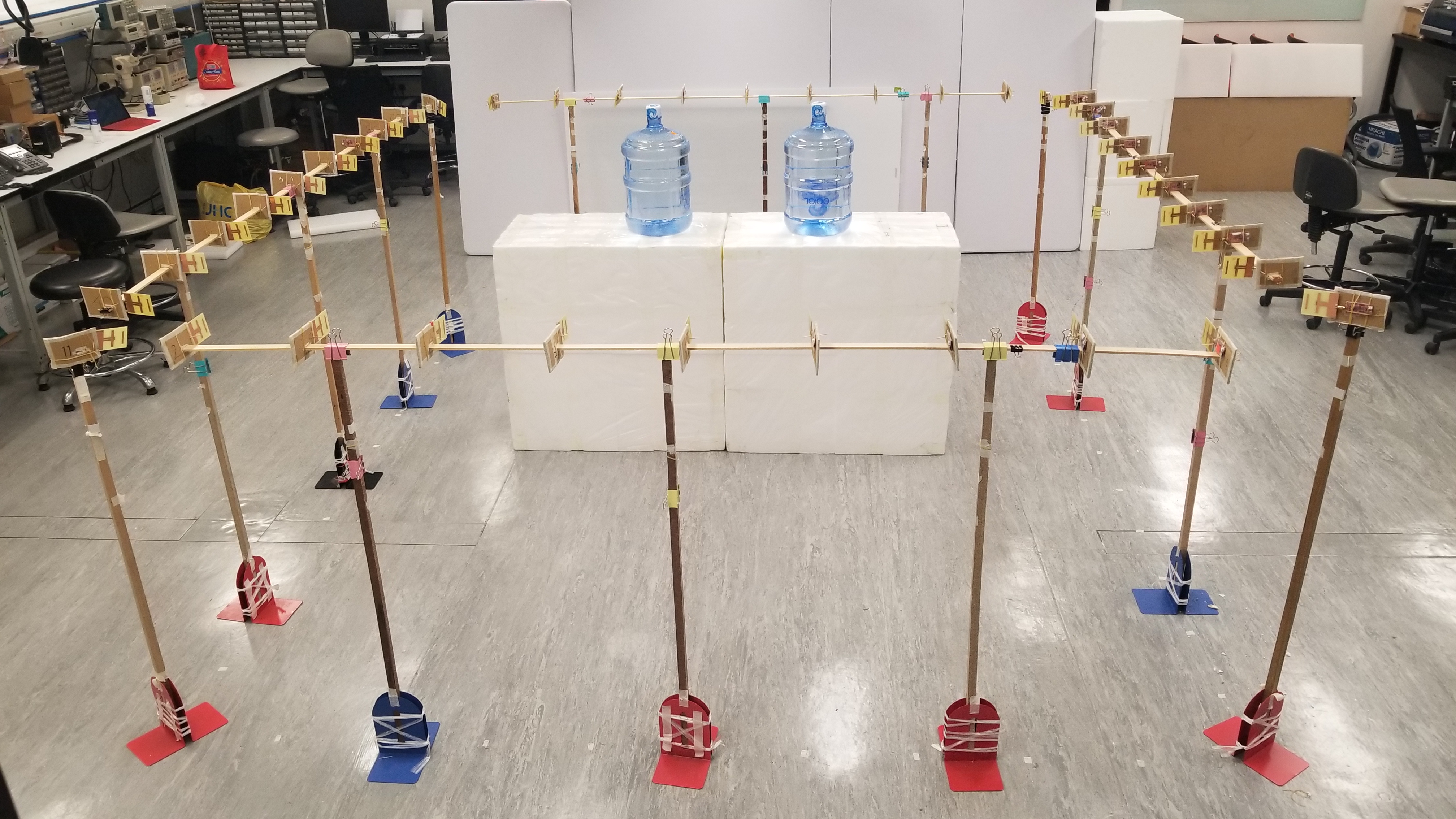}
		\includegraphics[scale=0.19]{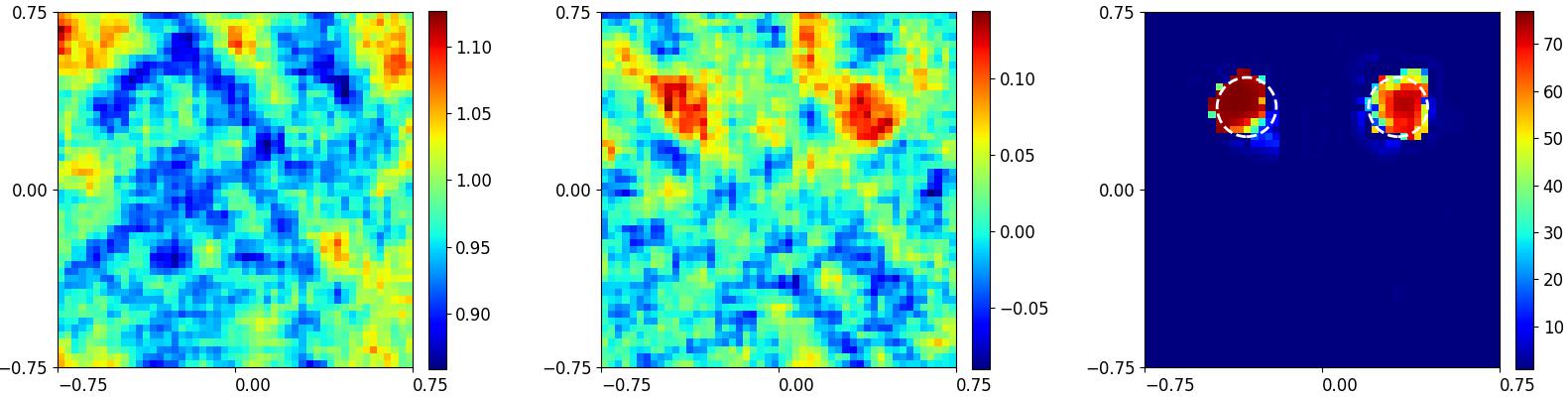}
		\subcaption{Two containers placed at a distance}
	\end{subfigure}
	\caption{Reconstruction results for one or more containers filled with water ($\epsilon_r = 77 + j7$). First column shows the experimental ground truth, the second and third columns show the real and imaginary parts of the xRA reconstruction respectively, and the last column shows the final $\epsilon_R$ reconstruction using our physics assisted deep learning approach. The respective PSNR values of $\epsilon_R$ reconstructions are (a) 26.25 db, (b) 20.10 db and (c) 17.44 db.}
	\label{Results_water}
\end{figure*}

\begin{figure*}
	\centering
	\includegraphics[scale=0.035]{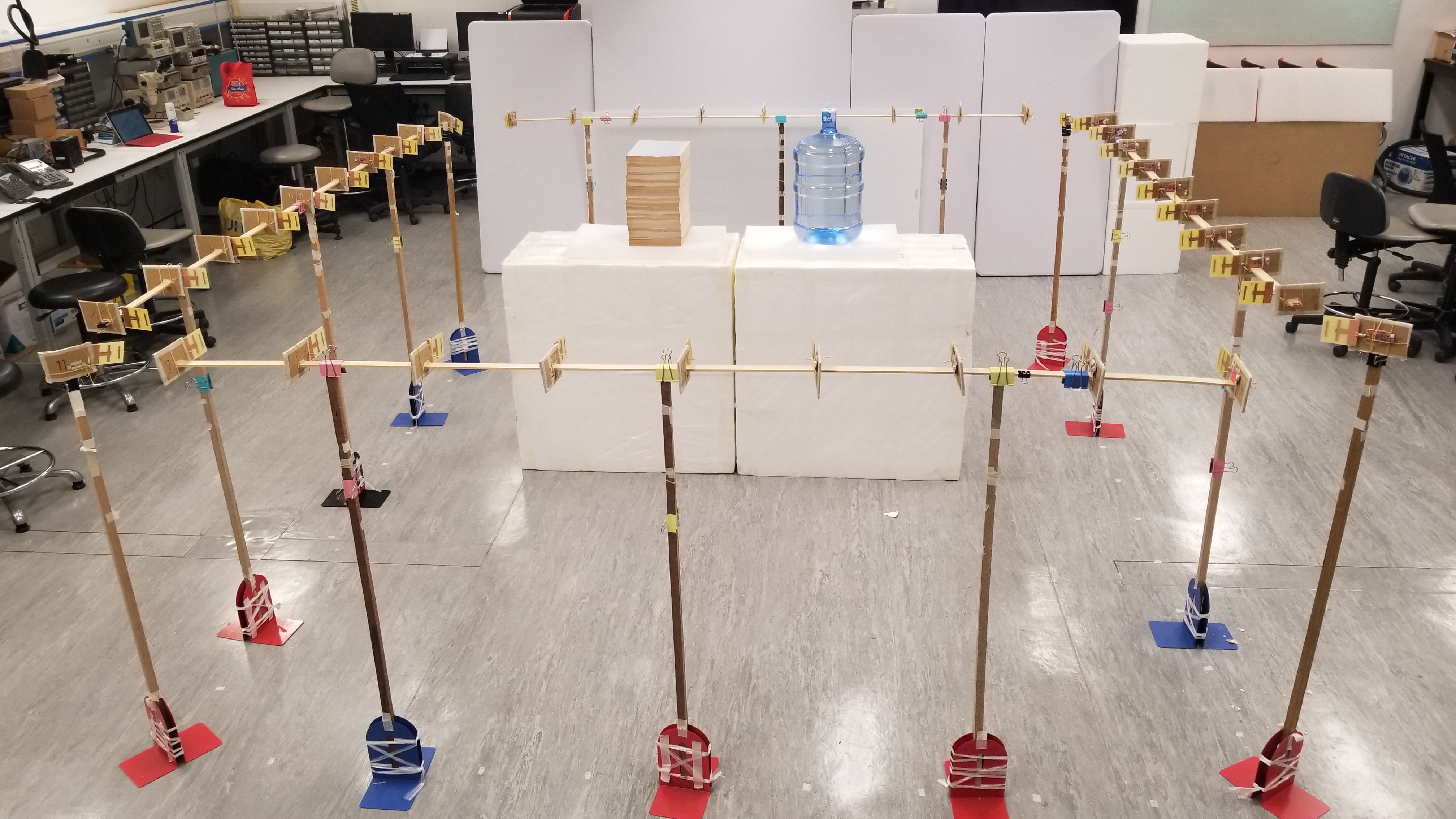}
	\includegraphics[scale=0.2]{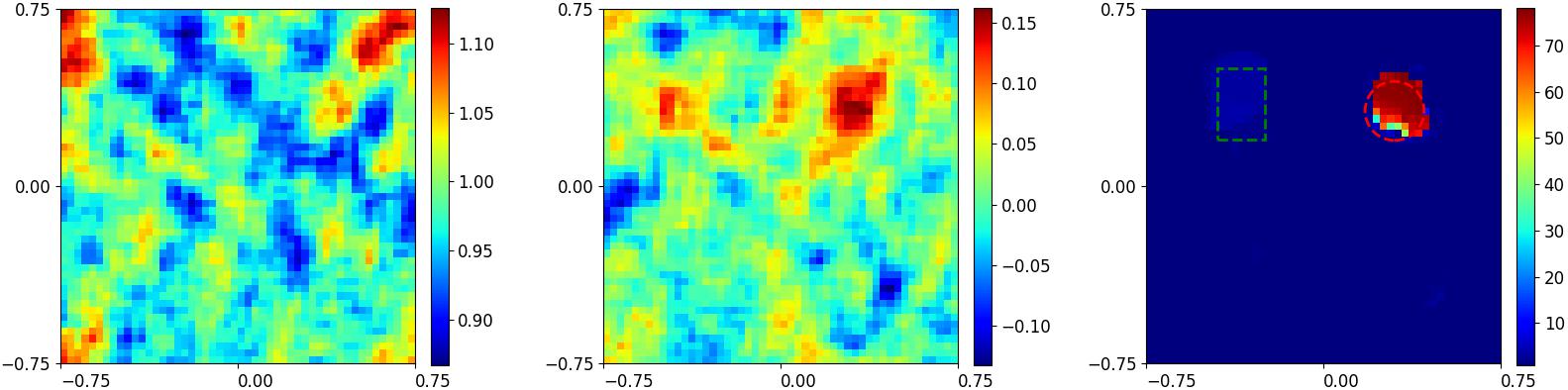}
	\caption{Reconstruction results for a stack of books and a container filled with water placed in the DoI together. First column shows the experimental ground truth, the second and third columns show the real and imaginary parts of the xRA reconstruction respectively, and the last column shows the final $\epsilon_R$ reconstruction using our physics assisted deep learning approach. The PSNR value for the reconstruction is 24.92 dB.}
	\label{Results_water_book}
\end{figure*}

\subsection{Experimental Tests}
We test our proposed framework on experimental data using the same modified U-Net model that is trained on simulation data. The most important point to note here is that the network is only trained only on simulation data for lossless scatterers ($\epsilon_r=\epsilon_R, \ \epsilon_I=0$), whereas in an actual indoor environment, almost all objects exhibit lossy behaviour ($\epsilon_r=\epsilon_R +j \epsilon_I$). For such cases the contrast in (\ref{Eq_newcontrast}) and the functions $f$ and $g$ in (\ref{Eq_newconstrastfinal}) are non-linear functions of both $\epsilon_R$ and $\epsilon_I$. Therefore, good experimental performance (as shown in this section) implies that the proposed framework is able to estimate $\epsilon_R$ profile even if (\ref{Eq_newcontrast}) is a function of both $\epsilon_R$ and $\epsilon_I$.

Figs. \ref{Results_books}, \ref{Results_water} and \ref{Results_water_book} show experimental results for different scatterers placed in the DoI where the PSNR reconstruction results are listed in the figure captions. In all these figures, the left-most columns show these scatterers placed on Styrofoam platforms in the experimental region. The second and third columns show the real and imaginary parts of the contrast ($\operatorname*{Re}(\chi_{\text{RI}})$ and $ \operatorname*{Im}(\chi_{\text{RI}})$) obtained using the xRA inverse model. The last column shows the final $\epsilon_R$ reconstruction of the scatterers, where the dashed line reveals the actual shape and location of scatterers in the DoI.

At an initial time $t_0$, we record RSSI measurements $P^{t_0}$ of the DoI containing Styrofoams but no scatterers. This data contains scattering from all objects inside the DoI including Styrofoams, floor, ceiling and other clutter present. We then place the scatterers on the Styrofoams at time $t_0 + \Delta t$ and take measurements $P^{t_0 + \Delta t}$. In addition to scattering due to other objects and clutter in the DoI, this measurement now also contains the effect of scatterers placed in the DoI. We then use the xRA inverse model to estimate the change in contrast profile due to the scatterer being placed. Passing this as input to the modified U-Net model gives the reconstruction of the scatterers placed at time $t_0 + \Delta t$, while ignoring the effect due to everything else in the DoI.

Fig. \ref{Results_books} shows experimental results when the scatterers are one or more stacks of books, each of which has dimensions 0.3$\times$0.2 m$^2$. For each of these book stacks, $\epsilon_r = 3.4 + j 0.25$ (as estimated using a high precision cavity resonator). As we can see, our proposed method provides accurate reconstruction of the scatterers using experimental data even when the network used was only trained on simulation data for lossless scatterers. There is additional distortion in the shape reconstruction of scatterers from experiments as compared to similar scatterers from simulation. This can be attributed to higher noise in RSSI measurements obtained from transceivers during experiments as well as the lossy nature of objects. 

Fig. \ref{Results_water} shows similar experimental results in which the scatterer considered is a cylindrical container of diameter 0.25 m filled with water which has a permittivity value of $\epsilon_r = 77 + j7$. Fig. \ref{Results_water}(a) shows a single water container being reconstructed along with its shape and permittivity value. Fig. \ref{Results_water}(b) and (c) each contain two such containers. 

Fig. \ref{Results_water_book} shows the result for a setup where both the book stack and the container filled with water are placed in the DoI. The proposed framework provides accurate reconstruction of both objects, and is able to provide a high dynamic range of permittivity values to clearly distinguish between the low permittivity stack of books and the high permittivity container of water. 

To summarize, our proposed method provides good results for experimental data obtained in a real indoor environment where the measurements are highly distorted (and different from simulation data used for training) due to multiple scattering from background clutter and lossy behavior of scatterers. The average PSNR value for the reconstructions is 21 dB. As far as we know, experimental results for accurately estimating the permittivity profile of actual indoor environments have not been demonstrated before. The ability of our proposed framework to generalize well to experimental data containing multipath reflections and clutter can be attributed to the background subtraction in the xRA inverse model.

\section{Conclusion}

We consider the problem of indoor RF imaging using phaseless Wi-Fi measurements to image objects with very high permittivity values and size comparable to or larger than the probing wavelength. A physics assisted deep learning approach is introduced which uses the xRA inverse model with temporal background subtraction to obtain an initial reconstruction profile. This model is used as input to a modified U-Net model to obtain the reconstruction of the real part of relative permittivity. Results for the proposed framework are demonstrated on simulation data as well as on data obtained from experiments in a real indoor environment containing multipath reflections and clutter. The results show accurate reconstruction of shape and permittivity values for different objects. The proposed framework exceeds the validity range of existing techniques by a significant margin and is able to provide accurate reconstructions for objects with relative permittivity as high as $77$ and size comparable to or larger than the incident wavelength.

Scope for future research includes reconstructing the imaginary component of the relative permittivity in addition to the real component. In addition reducing the number of measurements needed for the xRA reconstruction to make the system more robust for indoor imaging would be desirable. Other deep learning architectures that are more suitable for the inverse scattering use case can also be explored. Furthermore, reducing the amount of data needed to train the deep learning model to obtain robust reconstructions and incorporating model information in the framework to obtain general results for unseen environments and arbitrary object shapes are also useful goals.

\bibliographystyle{IEEEtran}
\bibliography{rytDL}


%

\end{document}